\documentclass[epj,onecolumn]{svjour}
\usepackage{graphicx}
\usepackage{lipsum}
\usepackage{float}
\usepackage{amsmath}
\usepackage{amssymb}
\usepackage{mathtools}
\include{mathmacros}

\begin{document}

\title{
Analytical results for 
the distribution of shortest path lengths
in directed random networks
that grow by node duplication
}

\author{Chanania Steinbock, Ofer Biham and Eytan Katzav}

\institute{
Racah Institute of Physics, 
The Hebrew University, 
Jerusalem 91904, Israel}

\authorrunning{Steinbock, Biham and Katzav}
\titlerunning{Shortest path lengths in directed node duplication networks}

\abstract{
We present exact analytical results for the 
distribution of shortest path lengths (DSPL)
in a directed network model that grows by node duplication.
Such models are useful in the study of the structure and growth dynamics of
gene regulatory networks and scientific citation networks.
Starting from an initial seed network, at each time step
a random node, referred to as a mother node, is selected for duplication.
Its daughter node is added to the network 
and duplicates each outgoing link of the mother node
with probability $p$.
In addition, the daughter node forms a directed link to the 
mother node itself. 
Thus, the model is referred to as the corded directed-node-duplication
(DND) model.
In this network not all pairs of nodes are connected by
directed paths, in spite of the fact that the corresponding undirected network
consists of a single connected component.
More specifically, in the large network limit only a diminishing fraction
of pairs of nodes are connected by directed paths.
To calculate the DSPL between those pairs of nodes that are
connected by directed paths we
derive a master equation for the time
evolution of the probability 
$P_t(L=\ell)$, $\ell=1,2,\dots$,
where $\ell$ is the length of the shortest directed path.
Solving the master equation, 
we obtain a closed form expression for 
$P_t(L=\ell)$.
It is found that the DSPL at time $t$ consists of a convolution of the
initial DSPL $P_0(L=\ell)$, with a Poisson distribution and a sum of Poisson distributions.
The mean distance 
${\mathbb E}_t[L|L<\infty]$
between pairs of nodes which are connected by 
directed paths
is found to depend logarithmically on the network size $N_t$.
However, since in the large network limit the fraction of pairs of nodes that are connected
by directed paths is diminishingly small, 
the corded DND network is not a small-world network,
unlike the corresponding undirected network.
}
\maketitle

\section{Introduction}

The increasing interest in the field of 
complex networks in recent years is motivated by 
the realization that a large variety of systems and processes
in physics, chemistry, biology, engineering, and society 
can be usefully described by network models
\cite{Albert2002,Caldarelli2007,Havlin2010,Newman2010,Estrada2011b,Barrat2012}.
These models consist of nodes and edges, where the nodes
represent physical objects, while the edges represent the
interactions between them.
A common feature of complex networks is the small-world property,
namely the fact that the mean distance and the diameter 
scale like $\ln N$,
where $N$ is the network size
\cite{Milgram1967,Watts1998,Chung2002,Chung2003}.
Many of these networks are scale-free,
which means that they exhibit power-law degree distributions
\cite{Barabasi1999,Jeong2000,Krapivsky2000,Krapivsky2001,Vazquez2003}.
The most highly connected nodes, called hubs, 
play a dominant role in dynamical processes on these networks.
Moreover, it was shown that scale-free networks are generically
ultrasmall, namely their mean distance and diameter scale like
$\ln \ln N$
\cite{Cohen2003}.

While pairs of adjacent nodes exhibit direct interactions,
the interactions between most pairs of nodes
are indirect, and are mediated by intermediate nodes and edges.
Pairs of nodes may be connected by many different paths. 
The shortest among these paths are of particular 
importance because they are likely to provide the fastest 
and strongest interactions.
Therefore, it is of much interest to study the 
distribution of shortest path lengths (DSPL) 
between pairs of nodes in different types of networks.
The DSPL is
of great importance for the 
temporal evolution of dynamical processes
\cite{Barrat2012}, 
such as signal propagation in genetic regulatory networks
\cite{Giot2003,Maayan2005}, 
navigation 
\cite{Dijkstra1959,Delling2009}
and epidemic spreading 
\cite{Satorras2015}.
Central measures of the DSPL such as 
the mean distance 
and extremal measures such as
the diameter 
were studied
\cite{Watts1998,Bollobas2001,Durrett2007,Fronczak2004,Newman2001b,Hartmann2017}.
However, apart from a few studies
\cite{Newman2001,Dorogotsev2003,Blondel2007,Hofstad2007,Esker2008,Shao2008,Shao2009},
the DSPL has not attracted nearly as much 
attention as the degree distribution.
Recently, an analytical approach was developed for calculating 
the DSPL 
\cite{Katzav2015}
in the
Erd{\H o}s-R\'enyi (ER) network
\cite{Erdos1959,Erdos1960,Erdos1961},
which is the simplest mathematical model of a random network.
More general formulations were later developed 
\cite{Nitzan2016,Melnik2016},
for the broader class of 
configuration model networks
\cite{Newman2001,Molloy1995,Molloy1998}. 

To gain insight into the structure of complex networks, 
it is useful to study the growth dynamics that gives rise to these structures.
In general, it appears that many of the networks encountered in biological,
ecological and social systems grow step by step, by the addition of new nodes
and their attachment to existing nodes. 
A common feature of these growth processes is
the preferential attachment mechanism, in which the likelihood of
an existing node to gain a link to the new node 
is proportional to its degree.
It was shown that growth models based on preferential
attachment give rise to scale-free networks,
which exhibit power-law degree distributions
\cite{Albert2002,Barabasi1999}. 
The effect of node duplication (ND) processes 
on network structure 
was studied using an undirected network growth model
in which at each time step a random node, 
referred to as a mother node,
is selected for duplication
and its daughter node duplicates
each link of the mother node with probability $p$
\cite{Bhan2002,Kim2002,Chung2003b,Krapivsky2005,Ispolatov2005,Ispolatov2005b,Bebek2006,Li2013}.
In this model the daughter node does not form a link
to the mother node, and thus in the following it
is referred to as the
uncorded ND model. 
It was shown that for $0 < p < 1/2$ 
the resulting network exhibits a power law
degree distribution of the form 

\begin{equation}
P(K=k) \sim k^{-\gamma}.
\label{eq:Pkkg}
\end{equation}

\noindent
For 
$0 < p < 1/e$,
where $e$ is the base of the natural logarithm,
the exponent is given by the nontrivial solution of
the equation
$\gamma = 3 - p^{\gamma-2}$, 
while for 
$1/e \le p < 1/2$ 
it takes the value 
$\gamma=2$
\cite{Ispolatov2005}.
For $1/2 \le p \le 1$ the degree distribution does not converge 
to an asymptotic form.

Recently, a different 
variant of an undirected
node duplication model
was introduced and studied
\cite{Lambiotte2016,Bhat2016}.
In this model, referred to as the corded ND model, 
at each time step a random mother node, M,
is selected for duplication.
The daughter node, D, is added to the network.
It forms an undirected link to its mother node, M,
and is also connected with probability $p$ to each 
neighbor of M.
It was shown that for 
$0 < p < 1/2$ 
the corded ND model generates a sparse network,
while for
$1/2 \le p \le 1$ the model gives rise to a dense network in 
which the mean degree increases with the network size
\cite{Lambiotte2016,Bhat2016}.
For $0 < p < 1/2$ 
the degree distribution 
of this network follows 
a power-law distribution,
given by Eq. (\ref{eq:Pkkg}),
where the exponent 
$\gamma$
is given by the non-trivial solution of the
equation 
$\gamma = 1 + p^{-1} - p^{\gamma-2}$
\cite{Lambiotte2016,Bhat2016}.
In the limit of $p \rightarrow 0$, the exponent 
diverges like
$\gamma \sim 1/p$.
This model is suitable for the description of acquaintance networks,
in which a newcomer who has a friend in the 
community becomes acquainted with other members 
\cite{Toivonen2009}.
Unlike the uncorded ND model, 
the formation of triadic
closures is built-in to the dynamics of the corded ND model. 
This means that once the
daughter node forms a link to a neighbor of the mother node, it completes
a triangle in which the mother, neighbor and daughter nodes are all connected 
to each other. The formation of triadic closures is an essential property of
the dynamics of social networks
\cite{Granovetter1973}. 
The formation of triadic closures is in sharp contrast to
configuration model networks, which exhibit a local
tree-like structure.
Interestingly, many empirical networks exhibit a high 
abundance of triangles,
both in undirected networks
\cite{Newman2001b}
and in directed networks,
where most triangles form feed-forward loops (FFLs), while triangular
feedback loops are rare
\cite{Milo2002,Alon2006}.
Unlike configuration model networks
\cite{Newman2001,Molloy1995,Molloy1998}, 
which may include small, isolated components,
the corded ND network consists of a single connected component.
Therefore, it does not exhibit a percolation transition.

In a recent paper, we studied the 
DSPL of the (undirected) corded ND network
\cite{Steinbock2017}.
Focusing on the
sparse network regime of $0 < p < 1/2$,
we derived a master equation for the
time evolution of the probabilities
$P_t(L=\ell)$,
where $\ell=1,2,\dots$ is the distance between a pair of nodes
and $t$ is the time.
Solving the master equation we
obtained an expression for
$P_t(L=\ell)$,
which consists of two convolution-like sums.
The first sum emanates from the
DSPL of the seed network
$P_0(L=\ell)$,
while the second sum involves a discrete 
exponential function.
We calculated the mean distance
$\langle L \rangle_t$,
and the diameter,
$\Delta_t$,
and showed
that in the long-time limit they scale like
$\ln t$,
namely the corded ND network is a small-world network
\cite{Milgram1967,Watts1998,Chung2002,Chung2003}.

In a more recent paper we introduced a directed version of the
corded node duplication network model and studied its in-degree
and out-degree distributions
\cite{Steinbock2018}. 
This model is referred to as the
corded directed node duplication (DND) model.
At each time step a random mother node is chosen for 
duplication. The daughter node forms a directed
link to the mother node and with probability $p$ to each outgoing
neighbor of the mother node.
This model may be useful in the study of
gene regulatory networks.
These are directed networks that evolve by gene duplication
\cite{Ohno1970,Teichmann2004}.
It also describes the structure and dynamics of scientific citation networks
\cite{Redner1998,Redner2005,Radicchi2008,Golosovsky2012,Golosovsky2017a,Golosovsky2017b},
in which the nodes represent papers, while the links
represent citations.
Scientific citation networks are
directed networks, with links pointing from the later (citing) paper
to the earlier (cited) paper.
A paper A, citing an earlier paper B,
often also cites one or several papers C,
which were cited in B
\cite{Peterson2010}.
The resulting network module is a triangle, or triadic closure,
which includes three directed links, from A to B, from B to C and
from A to C and thus resembles the FFL structure.
However, the links of this module point backwards, 
and thus it may be more suitable to refer to it as a 
feed-backward loop (FBL).
The functionality of FBLs is different from that of FFLs.
FFLs typically appear in information processing networks, enabling
the interference between signals propagating along two different paths.
In contrast, FBLs track the source of the cited information both directly and indirectly
via an intermediate reference point.
Due to the directionality of the links, each node exhibits both an in-degree,
which is the number of incoming links and an out-degree which is the number
of outgoing links.
Therefore, the degree distribution consists of two separate distributions,
namely the distribution $P_t(K_{\rm in}=k)$ of in-degrees and the distribution
$P_t(K_{\rm out}=k)$ of out-degrees, at time $t$. 
These two distributions are related to each other by the constraint that their means
must be equal, namely
$\langle K_{\rm in} \rangle_t = \langle K_{\rm out} \rangle_t$.
We obtained exact analytical results for 
the in-degree distribution and the out-degree distribution of the
corded DND network
\cite{Steinbock2018}.  
It was found that the in-degrees follow a shifted power-law distribution
while the out-degrees follow a narrow distribution  
that converges to a Poisson distribution in the sparse network limit
and to a Gaussian distribution in the dense network limit.
Since the network is directed not all pairs of nodes are connected by
directed paths even though the corresponding undirected network
consists of a single connected component.
We also obtained analytical results for the distribution
of the number of upstream nodes
$P_t(N_{\rm up}=n)$,
and for the distribution of the number of downstream nodes
$P_t(N_{\rm down}=n)$,
and show that 
$\langle N_{\rm up} \rangle = \langle N_{\rm down} \rangle$
are logarithmic in the network size.
This means that in the large network limit only a diminishing fraction
of pairs of nodes are connected by directed paths.

In this paper we present exact analytical results for the DSPL of the
corded DND network.
In order to calculate the DSPL we
derive a master equation for the time evolution of the
probability $P_t(L=\ell)$ that the shortest directed path from
a random node $i$ to a random node $j$ at time $t$ is of
length $\ell$, where the probability that there is no directed path
from $i$ to $j$ is given by $P_t(L=\infty)$.
Solving the master equation we
obtain a closed form analytical expression for the DSPL, which
consists of two terms.
The first term is a convolution of the initial DSPL $P_0(L=\ell)$, with a 
Poisson distribution, while the second term is
a sum of Poisson distributions.
The mean distance 
${\mathbb E}_t[L|L<\infty]$,
between pairs of nodes that are connected by
directed paths,
is found to be logarithmic in the network size $N_t$.
However, in the large network limit the fraction of pairs of nodes that are connected
by directed paths is diminishingly small.
Therefore, the corded DND network is not a small-world network,
unlike the corresponding undirected network.

The paper is organized as follows.
In Sec. 2 we present the corded DND model.
In Sec. 3 we consider the distribution of degeneracy levels,
$P(G=g)$, of the shortest paths between
random pairs of nodes.
In Sec. 4 we analyze the effect of these degeneracies on the
DSPL.
In Sec. 5 we provide a closed form analytical expression for   
$P_t(L=\ell)$, using a slightly approximated version of the master equation. 
The mean distance is calculated in Sec. 6 and the variance of the DSPL
is evaluated in Sec. 7. 
The results are discussed in Sec. 8 and summarized in Sec. 9.
In Appendix A we consider the set of canonical downstream 
configurations, which is used in the analysis of the degeneracies
of shortest paths.
In Appendix B we present the master equation for the temporal evolution
of the degeneracy distribution $P_t(G=g)$, truncated at $g=2$
and evaluate the rate of congergence to the asymptotic form.
In Appendix C we present the exact form of the master equation for the DSPL, 
and its time dependent solution, without the approximation used in Sec. 5.

\begin{figure}
\begin{center}
\includegraphics[width=7cm]{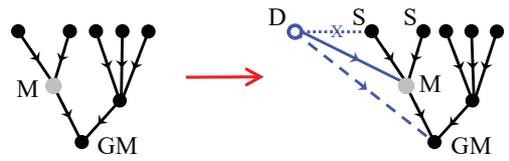}
\end{center}
\caption{
Illustration of the corded DND model.
A random node, referred to as a mother node, M (gray circle)
is selected for duplication. The newly formed daughter node, D
(empty circle) deterministically acquires a directed link (solid line) 
to the mother node. It also acquires, with probability $p$,
a directed link (dashed line) to each one of the outgoing neighbors of M.
In this example, D forms a directed link
to its grandmother node, denoted by GM.
The model does not allow D to form links to its sister nodes, denoted by S,
because they are incoming neighbors rather than outgoing neighbors of M.
}
\label{fig:1}
\end{figure}

\section{The corded directed node duplication model}

\begin{figure}
\begin{center}
\includegraphics[width=5.0cm]{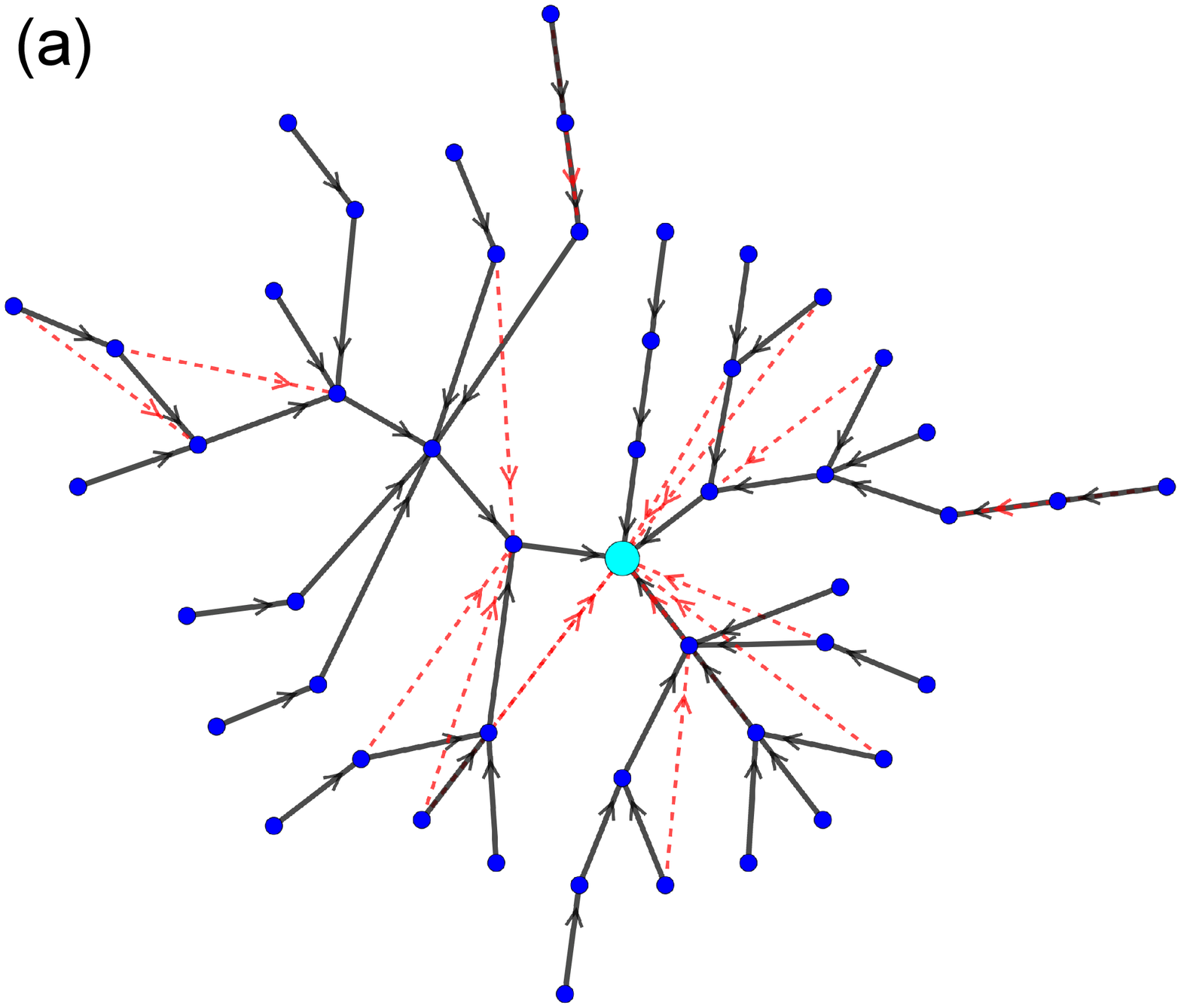} 
\vspace{0.5in}
\\
\includegraphics[width=5.0cm]{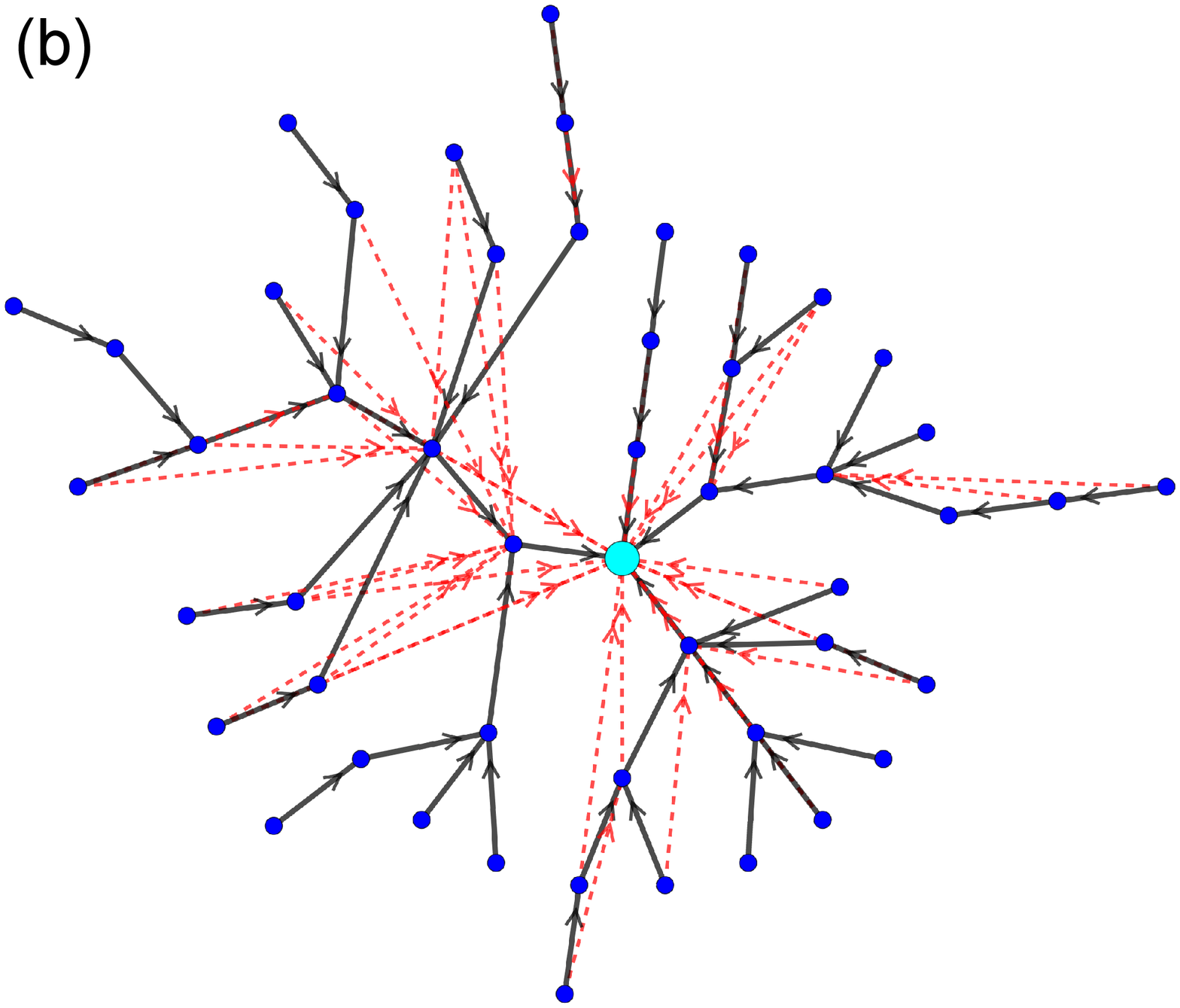}
\end{center}
\caption{
Two instances of corded DND networks of size $N=50$,
with $p=0.2$ (a) and $p=0.5$ (b).
Both networks were grown from a seed network that consists of
two nodes connected by a directed link.
For the sake of comparison, both instances are formed around
the same backbone tree (solid lines).
The sink node, which can be reached from all the other nodes in
the network via directed paths is shown by a large circle.
The probabilistic links (dashed lines) essentially decorate the
tree. Increasing $p$ makes the network denser.
}
\label{fig:2}
\end{figure}

In the corded DND model,
at each time step during the growth phase of the
network, a random node,
referred to as a mother node,
is selected for duplication.
The daughter node is added to the network, 
forming a directed link to the mother node.
Also, with probability $p$, it forms a directed link
to each outgoing neighbor
of the mother node (Fig. \ref{fig:1}).
The growth process starts from an initial 
seed network of $N_0=s$ nodes.
Thus, the network size
after $t$ time steps is 
$N_t = t + s$.
In Fig. \ref{fig:2}
we present two instances of the corded DND network, of size $N_t=50$,
which were formed around the same backbone tree.
Both networks were grown from a seed network of size $s=2$,
with $p=0.2$ [Fig. \ref{fig:2}(a)] and $p=0.5$ [Fig. \ref{fig:2}(b)].
Thus, each network instance includes $N_t-1=49$ deterministic links
(solid lines).
The network of Fig. \ref{fig:2}(b) is denser and includes $39$
probabilistic links (dashed lines), 
compared to $19$ probabilistic links in 
Fig. \ref{fig:2}(a).

Upon formation, the in-degree of the daughter node is zero. 
The incoming links are gradually formed as the daughter node matures.
Since the mother node at time $t$ is selected randomly from
all the $N_t$ nodes in the network, its in-degree
is effectively drawn from the in-degree distribution $P_t(K_{\rm in}=k)$.
The mother node gains an incoming link from the daughter node, thus
its in-degree increases by $1$.
The daughter node gains one outgoing link to the mother node,
and with probability $p$ it duplicates outgoing
links of the mother node.
Thus, in the case that all the outgoing links of the mother node are duplicated, 
the out-degree of the daughter node becomes
$k_{\rm out}^{\rm D} = k_{\rm out}^{\rm M} + 1$.
In the case that none of the outgoing links of the mother node
are duplicated the out-degree of the daughter node becomes
$k_{\rm out}^{\rm D}=1$.

In order to obtain a network that consists of a
single connected component, it is required
that the seed network consist of a single connected
component.
The size of the seed network is denoted by $s$. 
Since the corded DND model generates oriented networks with no bidirectional edges,
we restrict the analysis to seed networks in which a pair of nodes cannot
be connected in both directions. 
Moreover, we consider only acyclic seed networks, 
namely networks that do not include any
directed cycles.
Finally, we focus on seed networks that include a single
sink node, namely a node that has only incoming links and
no outgoing links.
The sink node can be reached via directed paths from all the nodes
in the seed network.
The in-degree distribution of the seed network is denoted by 
$P_0(K_{\rm in}=k)$
and its out-degree distribution is denoted by
$P_0(K_{\rm out}=k)$.
Clearly, the mean of the in-degree distribution
and the mean of the out-degree distribution
are equal to each other.
We thus denote
$\langle K \rangle_0 = \langle K_{\rm in} \rangle_0 = \langle K_{\rm out} \rangle_0$.

The DSPL of the seed network is denoted by
$P_0(L=\ell)$, $\ell =1,2,\dots,s-1$.
The probability that a random pair of nodes $i$ and $j$, 
in the seed network,
are connected by a directed path from $i$ to $j$ is
denoted by 

\begin{equation}
P_0(L < \infty) = \sum_{\ell=1}^{s-1} P_0(L=\ell).
\end{equation}

\noindent
The complementary probability, 
$P_0(L = \infty) = 1 - P_0(L < \infty)$, is the probability
that there is no directed path from $i$ to $j$.
The mean distance between directed pairs of nodes in the seed network 
that are connected by directed paths
is denoted by

\begin{equation}
\langle L \rangle_0 = \frac{ \sum\limits_{\ell=1}^{s-1} \ell P_0(L=\ell) }{ P_0(L < \infty) }.
\end{equation}

\noindent
The DSPL and the mean in-degree (or out-degree) of the seed network are related by
$P_0(L=1) = \langle K \rangle_0 / (s-1)$.
The probability $P_0(L=\ell)$ may take non-zero values
for $\ell=1,2,\dots,\Delta_0$,
where $\Delta_0$ is the diameter of the seed network,
while $P_0(L=\ell)=0$ for $\ell \ge \Delta_0 + 1$.
For seed networks of $s$ nodes, 
$\Delta_0$ may take values in the range
$1 \le \Delta_0 \le s-1$.

To avoid memory effects, which may slow down the
convergence to the asymptotic structure, it is often convenient to use
a seed network that consists of a single node,
namely $s=1$.
For a seed network that consists of a single node the
initial DSPL at $t=0$ is not defined.
However, 
the DSPL becomes well defined
at time $t=1$, when the network consists of
a pair of connected nodes,
whose 
DSPL is given by
$P_1(L=\ell) = \delta_{\ell,1}/2$,
where $\delta_{\ell,\ell'}$ is the Kronecker delta,
and
$P_1(L=\infty) = 1/2$,
while its diameter is 
$\Delta_1=1$.
Another interesting choice for the seed network is a
linear chain of $s$ nodes, in which all the links are
in the same direction.
In this case, the 
initial DSPL is

\begin{equation}
P_0(L=\ell) = \frac{s-\ell}{
s(s-1)
},
\end{equation} 

\noindent
for $\ell=1,2,\dots,s-1$
and $P_0(L=\infty)=1/2$.
This choice captures the largest possible diameter
in a seed network of $s$ nodes, namely 
$\Delta_0=s-1$.
For simplicity, all the analytical results and the corresponding
simulation results presented in the Figures below, were obtained
for a seed network that consists of a linear chain with $s=2$, 
namely a pair of nodes with a single directed link between them.

The mother-daughter links in the
corded DND network form a random directed tree structure,
which serves as a backbone tree for the resulting network.
The backbone tree is a random directed recursive tree
\cite{Smythe1995,Drmota1997,Drmota2005}.
To study its properties, one can take the limit of $p=0$,
in which the corded DND network is reduced to the backbone tree.

\section{The degeneracy of the shortest paths}

Consider a pair of nodes $i$ and $j$ for which there is at least one directed
path from $i$ to $j$. In the case that the shortest path from $i$ to $j$ is of
length $\ell \ge 2$, this path
may be unique or it may be degenerate. 
In the case that the shortest path is degenerate,
there are at least two different paths of
length $\ell$ from $i$ to $j$ (which may have overlapping segments).
In particular, the degenerate paths may differ in the first step, starting
from node $i$.
Here we focus on the degeneracy of the first step, namely on the
number of outgoing neighbors of $i$ which reside on shortest paths
from $i$ to $j$. 
The number of such distinct neighbors is denoted by $g$.
We denote the distribution of the degeneracy levels of
the first steps of the shortest paths by $P(G=g)$, where $g=1,2,\dots$.
In order to calculate the distribution $P(G=g)$ we follow the growth 
process of the network and consider the shortest path from the newly
formed daughter node, D, to a randomly selected node T downstream of M. 
It is important to note that the distances $\ell_{\rm DT}$ between the daughter node, D, and
all the accessible nodes, T, downstream of M are determined upon formation
of the node D. This is due to the fact that nodes and edges which will be added later cannot 
form paths between D and T which are shorter than $\ell_{\rm DT}$.
Moreover, unlike the case of the  corded undirected ND model, 
they cannot even form additional paths of length $\ell_{\rm DT}$
between D and T. This means that the degeneracies of the shortest
paths from D to all downstream nodes are also determined upon formation
of D.

\begin{figure}
\begin{center}
\includegraphics[width=7cm]{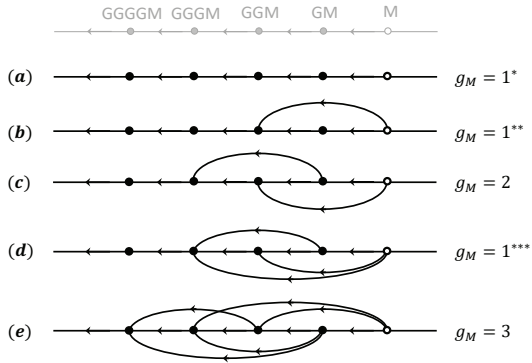}
\end{center}
\caption{
Illustration of possible network structures in the vicinity of 
a newly formed daughter node D and its mother node M.
The grandmother (GM) node, the great-grandmother (GGM) node
and two more generations are also marked downstream of M.
The deterministic links are represented by straight horizontal arrows
while the probabilistic links are represented by arcs.
(a) A linear branch of the backbone tree (marked by $g_{\rm M}=1^{*}$),
consisting of a succession of mother-daughter pairs connected
by deterministic links and no probabilistic links.
The nodes which are accessible from M reside downstream,
on the left hand side. The degeneracy of
the first step in the paths from M to all the nodes
that are accessible from M along directed paths is $g=1$;
(b) A configuration in which M forms a probabilistic path to GGM
(marked by $g_{\rm M}=1^{**}$).
The degeneracies of the downsteam paths from M are $g=1$.
However, this configuration differs from the configuration 
shown in (a) in the sense that in a future step, when
M will be selected for duplication, its daughter node D
may acquire degeneracy $g=2$ by forming a probabilistic
link to GM;
(c) A configuration in which the degeneracy of the first step in the paths from M to GGGM
and all downstream nodes is $g=2$ (marked by $g_{\rm M}=2$).
(d) In this configuration (marked by $g_{\rm M}=1^{***}$), the degeneracy of the paths downstream
of M is $g=1$. However, upon future duplication of M, its daughter node D
may acquire a degeneracy of $g=3$ 
for their paths to GGGM and all downstream nodes,
by forming probabilistic links to GM and GGM.
(e) A configuration in which the degeneracy of the first step in the paths from M
to GGGM and all its downstream nodes is $g=3$ (marked by $g_{\rm M}=3$).
}
\label{fig:3}
\end{figure}

In order to analyze the distribution of degeneracies, $P(G=g)$ we will
first classify all the possible local configurations downstream of a random
node $i$. These configurations determine the degeneracy of the paths from
$i$ to all its downstream nodes.
In Fig. \ref{fig:3} we present an 
illustration of the possible network structures in the downstream vicinity of 
a random node upon its selection as a mother node, M.
The grandmother (GM) node and the great-grandmother (GGM) node,
as well as two earlier generations (GGGM and GGGGM) 
are also marked downstream of M.
The deterministic links are represented by straight horizontal arrows,
pointing to the left.
The probabilistic links are represented by arcs.
More specifically, in Fig. \ref{fig:3}(a) we present
a linear branch of the backbone tree (marked by $g_{\rm M}=1^{*}$),
consisting of a succession of mother-daughter pairs connected
by deterministic links and no probabilistic links.
The nodes which are accessible from M reside downstream,
on the left hand side. The degeneracy of
the first step in the paths from M to all the nodes
which are accessible from M along directed paths is $g=1$;
In Fig. \ref{fig:3}(b) we present a
configuration in which M forms a probabilistic path to GGM.
The degeneracies of the directed paths from M to downsteam nodes is $g=1$.
However, in this configuration (marked by $g_{\rm M}=1^{**}$), 
upon duplication of M, its daughter node, D,
may acquire degeneracy $g=2$ by forming a probabilistic
link to GM.
In Fig. \ref{fig:3}(c) we present a
configuration in which the degeneracy of the first step in the paths from M to GGGM
and all downstream nodes is $g=2$ (marked by $g_{\rm M}=2$).
This is due to the fact that there are two different paths of length $\ell=2$
from M to GGGM, one path via GM and the other via GGM.
In Fig. \ref{fig:3}(d) we
present a configuration (marked by $g_{\rm M}=1^{***}$), 
in which the degeneracy of the shortest paths to all the nodes 
that reside downstream of M is $g=1$. 
However, upon duplication of M, its daughter node D
may acquire degeneracy of $g=3$ 
for the path to GGGM and all downstream nodes.
In order to acquire such degeneracy, 
D should duplicate the links from M to GM and to GGM, but should not duplicate the link to GGGM.
In general, the notation 
$g_{\rm M}=1^{**\dots*}$
represents configurations in which the degeneracy of the paths downstream of the
mother node is $g=1$, while the number of stars represents
the highest possible degeneracy of the paths downstream of the daughter node.
In Fig. \ref{fig:3}(e) we present a configuration in which the degeneracy 
of the shortest paths from M
to GGGGM and all its downstream nodes is $g=3$ (marked by $g_{\rm M}=3$).
This is due to the fact that there are three degenerate shortest paths of 
length $\ell=2$ from M to GGGGM: a path via GM, another path via GGM 
and a third path via GGGM.

\begin{figure}
\begin{center}
\includegraphics[width=7cm]{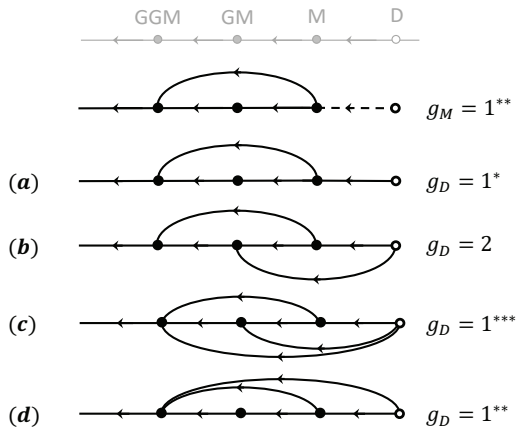}
\end{center}
\caption{
The evolution of the local downstream configuration and the
degeneracies of shortest paths,
from a mother node, M, to its daughter node D upon 
duplication of M. 
Top line: a random node, denoted by M, is selected for
duplication. The deterministic link from the daughter node D
to M is shown (dashed line). 
The node D may form probabilistic links to the two outgoing
neighbors of M, referred to as GM and GGM. 
Each one of the two links is formed with probability $p$,
and thus there are four possible downstream configurations for D.
(a) None of the links are formed and the resulting degeneracy
is $g=1$ (Configuration $g_{\rm D}=1^{*}$);
(b) Only the link from D to GM is formed, giving rise to $g=2$ (Configuration $g_{\rm D}=2$);
(c) Both links are formed and the resulting degeneracy is $g=1$
(configuration $g_{\rm D} =1^{***}$);
(d) Only the link to GGM is formed and the resulting degeneracy is $g=1$
(configuration $g_{\rm D}=1^{**}$).
}
\label{fig:4}
\end{figure}

In Fig. \ref{fig:4} we present 
the evolution of the local downstream configuration and the
degeneracies of shortest paths,
from a mother node M to its daughter node D, upon 
duplication of M. 
In the top line we show a random node, denoted by M, upon
its selection for duplication. 
The deterministic link from the daughter node D
to M is presented by a dashed line. 
The node D may form probabilistic links to the two outgoing
neighbors of M, namely GM and GGM. 
Each one of the two links is formed with probability $p$,
and thus there are four possible downstream configurations for D.
In Fig \ref{fig:4}(a) we present the case in which no probabilistic
links were formed. The resulting degeneracy of the paths from D 
to all the downstream nodes is $g=1$.
Therefore, with respect to the shortest paths and their degeneracies,
this configuration is equivalent to $g_{\rm D}=1^{*}$ (Appendix A).
In Fig. \ref{fig:4}(b) we present the case in which
the link to GM was duplicated, but the link to GGM was
not duplicated. In this case the degeneracies of the shortest
paths from D to GGM and all its downstream nodes is $g=2$ ($g_D=2$).
In Fig. \ref{fig:4}(c) we present the case in which the probabilistic links
to both GM and GGM are formed. 
The resulting configuration is $g_{\rm D}=1^{***}$,
in which the degeneracies of the shortest paths from D to all its 
downstream nodes are $g=1$,
but its future offsprings may acquire a double or triple degeneracy.
In Fig. \ref{fig:4}(d) we show the case in which the link to GGM
is duplicated, while the link to GM is not duplicated.
In this case the degeneracies of the shortest paths from D
to its downstream nodes is $g=1$
This configuration is equivalent to $g_D=1^{**}$ (Appendix A).

The probability distribution of the degeneracies of a randomly 
selected mother node, M, at time $t$,
is given a vector of the form
$\vec P_t^{\rm M} (G=g)$.
The elements of this vector represent the probabilities of
the downstream configurations of M,
given by 
$g=1^{*},1^{**},2,1^{***},3,\dots,g_{\rm max}$,
where 
$g_{\rm max}$ is
the truncation degeneracy level.
The transition probability from a configuration $g_{\rm M}$,
of the mother node, 
to a configuration $g_{\rm D}$ of its daughter node,
is denoted by 
$P(g_{\rm M} \rightarrow g_{\rm D})$. 
We present these transition probabilities in a matrix $T$, 
such that

\begin{equation}
\vec P_t^{\rm D} (G=g) = T \vec P_t^{\rm M}(G=g).
\end{equation}

\noindent
To demonstrate the dynamical evolution of the degeneracy we first
consider a truncated set of equations, which includes only three configurations:
$g=1^{*}$, $g=1^{**}$ and $g=2$.
In this case, the probability distribution of the degeneracies of a randomly 
selected mother node M, at time $t$,
is given by the vector

\begin{equation}
\vec P_t^{\rm M} (G=g) = 
\left[
\begin{array}{l}
P_t^{\rm M}(G=1^{*}) \\
P_t^{\rm M}(G=1^{**}) \\
P_t^{\rm M}(G=2) 
\end{array}
\right].
\label{eq:3st}
\end{equation}

\noindent
The probability distribution of the daughter node is given by
the corresponding vector $\vec P_t^{\rm D} (G=g)$.
The $3 \times 3$ transition matrix takes the form

\begin{equation}
T =
\left[ \begin{array}{ccc}
1-p \ \ & (1-p)^2 \ \  & (1-p)^2  \\
p \ \  & p \ \  & 2 p(1-p)  \\
0 \ \  & p(1-p) \ \ & p^2   
\end{array} \right].
\label{eq:T33}
\end{equation}

\noindent
Note that this matrix satisfies the condition
$\sum_{i=1}^{3} T_{i,j} = 1$
for $j=1$, $2$ and $3$, so probability is conserved.
In order to truncate the equations, we use a closure 
condition in which the transition probability
$T_{4,2} = P(1^{**} \rightarrow 1^{***})=p^2$
is added to
$T_{2,2} = P(1^{**} \rightarrow 1^{**})$.
Solving for the steady state of the equation
$\vec P(G=g) = T \vec P(G=g)$,
under the condition that
$P(G=1^{*})+P(G=1^{**})+P(G=2)=1$,
we obtain

\begin{eqnarray}
P(G=1^{*}) &=& \frac{(1-p)^2}{1-p+p^2}
\nonumber \\
P(G=1^{**}) &=& \frac{p(1+p)}{(1+2p)(1-p+p^2)}
\nonumber \\
P(G=2) &=& \frac{p^2}{(1+2p)(1-p+p^2)}.
\end{eqnarray}

\noindent
Summing up the contributions to $g=1$, using
$P(G=1)=P(G=1^{*}) + P(G=1^{**})$,
we obtain

\begin{eqnarray}
P(G=1) &=& \frac{1+p-2p^2+2p^3}{(1+2p)(1-p+p^2)}
\nonumber \\
P(G=2) &=& \frac{p^2}{(1+2p)(1-p+p^2)}.
\label{eq:PG12}
\end{eqnarray}

Extending the analysis up to degeneracy $g=3$,
the probability distribution of the degeneracies 
of all the nodes in the network at time $t$ is described
by the vector

\begin{equation}
\vec P_t(G=g) = 
\left[
\begin{array}{l}
P_t(G=1^{*}) \\
P_t(G=1^{**}) \\
P_t(G=2) \\
P_t(G=1^{***}) \\
P_t(G=3)
\end{array}
\right],
\label{P5}
\end{equation}

\noindent
and the transition matrix takes the form

\begin{eqnarray}
&&T =  
\nonumber \\
&& 
\left[ \begin{array}{ccccc}
1-p   & (1-p)^2 \  & (1-p)^2 \ & (1-p)^3 \   & (1-p)^3 \\
p &   p(1-p) \  &   2p(1-p)       \   & p(1-p)^2    \   &  3p(1-p)^2 \\
0 &   p(1-p) \  &   p^2         \   & 2p(1-p)^2   \    & 3p^2(1-p) \\
0 &   p^2    \ & 0                 \  &    p^2(2-p)  \   & 0 \\
0 &   0      \ & 0                   \    &   p^2(1-p)   \   &  p^3
\end{array} \right].
\nonumber \\
\label{eq:T55}
\end{eqnarray}

\noindent
Solving for the asymptotic steady state 
solution of the degeneracy vector,
given by
$\vec P(G=g) = T \vec P(G=g)$,
we obtain

\begin{eqnarray}
P(G=1^{*}) = \frac{(1-p)^2 (1+3p+2p^2 - 2p^4-2 p^5 +p^6)}
{1+2p+p^3+p^4+2p^5+4p^6 - p^7 + p^8} 
\nonumber \\
P(G=1^{**}) = \frac{p(1 + 2 p - 2 p^3 - 2 p^4  + p^6) }
{1+2p+p^3+p^4+2p^5+4p^6 - p^7 + p^8} 
\nonumber \\
P(G=2) = \frac{p^2(1+p+ p^2 - p^3 - p^4 + 2 p^5) } 
{1+2p+p^3+p^4+2p^5+4p^6 - p^7 + p^8} 
\nonumber \\
P(G=1^{***}) = \frac{p^3(1 + 2p + 2p^2 + p^3)  }
{1+2p+p^3+p^4+2p^5+4p^6 - p^7 + p^8} 
\nonumber \\
P(G=3) = \frac{p^5 (1+p)}
{1+2p+p^3+p^4+2p^5+4p^6 - p^7 + p^8}.
\nonumber \\
\label{eq:vecp}
\end{eqnarray}

\noindent
Summing up the three components which correspond to 
single degeneracy,
we obtain the probability 
$P(G=1) = P(G=1^{*}) + P(G=1^{**}) + P(G=1^{***})$.
The distribution of degeneracy levels at steady state is thus given by

\begin{eqnarray}
P(G=1) = \frac{1+2p-p^2+2p^5+4p^6-3p^7+p^8}
{1+2p+p^3+p^4+2p^5+4p^6 - p^7 + p^8} 
\nonumber \\
P(G=2) = \frac{p^2(1+p+ p^2 - p^3 - p^4 + 2 p^5) } 
{1+2p+p^3+p^4+2p^5+4p^6 - p^7 + p^8} 
\nonumber \\
P(G=3) = \frac{p^5 (1+p)}
{1+2p+p^3+p^4+2p^5+4p^6 - p^7 + p^8}.
\nonumber \\
\label{eq:vecp2}
\end{eqnarray}

\noindent
One can easily confirm that the probabilities in 
Eq. (\ref{eq:vecp2})
satisfy the normalization condition, namely
$P(G=1)+P(G=2)+P(G=3)=1$.
In the limit of $p \ll 1$ Eq. (\ref{eq:vecp2}) can be 
approximated by

\begin{eqnarray}
P(G=1) &=& 1 - p^2(1-p) + O(p^4)
\nonumber \\
P(G=2) &=& p^2(1-p) + O(p^4)
\nonumber \\
P(G=3) &=& p^5(1-p) + O(p^7).
\end{eqnarray}

\noindent
This reflects the fact that the double degeneracy requires two 
probabilistic links while the triple degeneracy requires five probabilistic links.
In both cases in order to maintain the degeneracy, the probabilistic link
that would shorten the degenerate paths and remove the degeneracy should not form.
This occurs with probability $1-p$.

\section{The effect of the degeneracy on the DSPL}

A node which resides at distance $\ell$ downstream of the mother node,
may end up either at distance $\ell$ or at distance $\ell+1$ 
from the daughter node.
To exemplify this property, consider a target node T at distance $\ell$
from the mother node M. 
A shortest path from M to T consists of a 
set of nodes
${\rm M},r_1,r_2,\dots,r_{\ell-1},{\rm T}$
in which subsequent nodes are connected by directed links.
In the case that the link between M and $r_1$ is duplicated by D,
the node T ends up at a distance $\ell$ from D,
while in the case it is not duplicated, the node T ends up at distance $\ell+1$
from D.
In the case that there is a single shortest path from M to T, the
former scenario would occur with probability $p$ while the latter
scenario would occur with probability $1-p$,
namely

\begin{equation}
P_t^{\rm D}(L=\ell) = p P_t^{\rm M}(L=\ell) +
(1-p) P_t^{\rm M}(L=\ell-1),
\end{equation}

\noindent
where $\ell \ge 2$.
However, since shortest paths of lengths $\ell \ge 2$ from M to T may be degenerate,
the calculation of 
$P_t^{\rm D}(L=\ell)$ 
requires a more careful attention.
For $\ell \ge 3$ we express the DSPL between the daughter node D and the nodes
that reside downstream of M in the form

\begin{equation}
P_t^{\rm D}(L=\ell) = \eta P_t^{\rm M}(L=\ell) + (1 - \eta) P_t^{\rm M}(L=\ell-1).
\label{eq:Pdell}
\end{equation}

\noindent
where 
$0< \eta <1$.
The case of $\ell=2$ is special in the sense that it combines the parameters
$p$ and $\eta$. It takes the form
 
\begin{equation}
P_t^{\rm D}(L=2) = \eta P_t^{\rm M}(L=2) + (1 - p) P_t^{\rm M}(L=1).
\label{eq:PdellLeq2}
\end{equation}

\noindent
Apart from the special treatment of paths of lengths $1$ and $2$, 
it is assumed that
$\eta = \eta(p)$ does not depend on the path length $\ell$.

In order to evaluate the parameter $\eta$,
consider a random target node T, 
which is at distance $\ell$ from the mother node M.
Here we are concerned with 
the degeneracy of the first step
along the shortest paths. 
This degeneracy is given by
the number of nearest 
neighbors of M which reside on at least
one shortest path from M to T,
and is denoted by
$g$.
Clearly, $g \le k$, where $k$ is the degree of the mother
node, M.
In the case that node M is chosen for duplication,
if none of the $g$ links of M 
which reside on shortest paths to T
are duplicated,
the distance between the daughter node 
D
and T 
becomes $\ell+1$, while in the case that at least
one of these $g$ edges is duplicated, the distance
is $\ell$.
Since each link of the mother node M is duplicated
with probability $p$, the probability that none of them
is duplicated is $(1-p)^g$.
The probability
that at least one of these $g$ links will be duplicated is 
$1-(1-p)^g$.
Thus, the probability $\eta$ that
at least one of the $g$ neighbors of the mother node M,
which reside along shortest paths to T, are 
connected to the daughter node can be expressed by

\begin{equation}
\eta = 1 -  \sum_{g=1}^{\infty}  (1-p)^{g} P(G=g).
\label{eq:eta}
\end{equation}

\noindent
Inserting $P(G=g)$ for $g=1$ and $2$ 
from Eq. (\ref{eq:PG12})
into Eq. (\ref{eq:eta}), 
we obtain

\begin{equation}
\eta = p + \frac{p^3(1-p)}{1+p-p^2+2p^3}.
\label{eq:eta12}
\end{equation}

\noindent
Inserting $P(G=g)$ for $g=1$, $2$ and $3$,
from Eq. (\ref{eq:vecp2})
into Eq. (\ref{eq:eta}), 
we obtain

\begin{equation}
\eta = p + \frac{(1-p) p^3 \left( 1+p+p^2+p^3+p^5 \right)}
{1+2p+p^3+p^4+2p^5+4p^6-p^7+p^8}.
\label{eq:eta5}
\end{equation}

\begin{figure}
\begin{center}
\includegraphics[width=7.0cm]{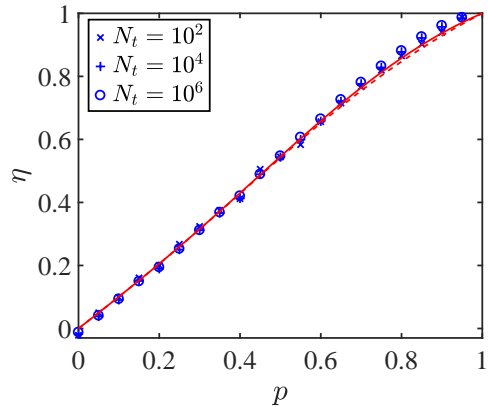}
\caption{
The probability $\eta$ 
that the shortest path length from the daughter node D to any downstream node T
is equal to the shortest path length from its mother node M to T,
as a function of $p$.
The analytical results obtained from
Eq. (\ref{eq:eta12}),
which takes into account only single and double degeneracies,
are shown by a dashed line,
while the results obtained from
Eq. (\ref{eq:eta5}),
which takes into account single, double and triple degeneracies,
are shown by a solid line.
The difference between the two curves is very small, indicating that the
results are well converged and that the effect of higher order degeneracies
is negligible.
For small values of $p$, where the shortest paths are non-degenerate,
$\eta$ is essentially equal to $p$. 
As $p$ is increased, the shortest paths are more likely to be degenerate. 
As a result, the probability $\eta$ becomes larger than $p$.
In the limit of $p \rightarrow 1$ the two probabilities converge.
The analytical results are found to be in very good agreement with
the results obtained from computer simulations (symbols) for
network sizes of $N_t=10^2$, $10^4$ and $10^6$.
}
\label{fig:5}
\end{center}
\end{figure}

In Fig. \ref{fig:5} we present the parameter $\eta$
as a function of $p$. 
The analytical results obtained from
Eq. (\ref{eq:eta12}),
which takes into account only single and double degeneracies,
are shown by a dashed line.
The results obtained from
Eq. (\ref{eq:eta5}),
which takes into account single, double and triple degeneracies,
are shown by a solid line.
The difference between the two curves is very small, indicating that the
results are well converged and that the effect of higher order degeneracies
is negligible.
For small values of $p$, where most of the shortest paths are non-degenerate,
$\eta$ is essentially equal to $p$. 
As $p$ is increased, the shortest paths are more likely to be degenerate. 
As a result, the probability $\eta$ becomes larger than $p$.
In the limit of $p \rightarrow 1$ the two probabilities converge.
The analytical results are found to be in very good agreement with
the results obtained from computer simulations (symbols) for
network sizes of $N_t=10^2$ ($\times$), $10^4$ ($+$) and $10^6$ ($\circ$).

\section{The distribution of shortest path lengths}

Consider an instance of the corded DND network.
At time $t$ there are $N_t(N_t-1)$ directed pairs of nodes,
in the network.
The probability that the shortest directed path from a random node, $i$,
to another random node $j$ is of length $\ell$ is denoted by
$P_t(L=\ell)$,
while the probability that there is no directed path from $i$ to $j$
is denoted by $P_t(L=\infty)$.
The longest possible path in a network of $N_t$ nodes is of length 
$\ell_{\rm max}(t) = N_t-1$. 
Thus, the distribution $P_t(L=\ell)$ satisfies the normalization
condition

\begin{equation}
\sum_{\ell=1}^{\ell_{\rm max}(t)} P_t(L=\ell) + P_t(L=\infty) = 1.
\end{equation}

\noindent
In the analysis below it will be convenient to define

\begin{equation}
P_t(L < \infty) = 1 - P_t(L = \infty).
\end{equation}

The probability that a randomly selected node cannot be
reached via a directed path from M is denoted by
$P_t^{\rm M}(L=\infty)$,
while the probability that such a node cannot be reached via a 
directed path from D is denoted by
$P_t^{\rm D}(L=\infty)$.
Since the daughter node forms a directed link to the mother node,
$P_t^{\rm D}(L=\infty)=P_t^{\rm M}(L=\infty)$.
Since upon formation, D does not have any incoming links, it cannot
be reached via a directed path from any one of the existing $N_t$ nodes
in the network.
After the incorporation of D into the network, the probability 
$P_{t+1}(L=\infty)$ 
is given by

\begin{eqnarray}
P_{t+1}(L=\infty) 
&=& 
\frac{  N_t(N_t-1) P_t(L=\infty) + N_t P_t^{\rm D}(L=\infty) }
{N_t(N_t+1)} 
\nonumber \\
&+& \frac{1}{N_t+1} 
- \frac{1}{(N_t+1)^2},
\label{eq:Ptinf0}
\end{eqnarray}

\noindent
where the last term accounts for the dilution of 
$P_{t+1}(L=\infty)$ by the deterministic link from D to M.
Replacing $P_t^{\rm D}(L=\infty)$  
by $P_t^{\rm M}(L=\infty)$
and using the fact that
the mother node is randomly selected at time $t$, 
namely
$P_t^{\rm M}(L=\infty) = P_t(L=\infty)$,
and combining the last two terms on the right hand side of Eq. (\ref{eq:Ptinf0}),
we obtain

\begin{equation}
P_{t+1}(L=\infty) = 
\frac{N_t}{N_t+1} P_t(L=\infty) + \frac{N_t}{(N_t+1)^2}.
\end{equation}

\noindent
Subtracting $P_t(L=\infty)$ from both sides,
expressing the difference on the left hand side as a derivative
and using the relation $N_t=t+s$,
we obtain

\begin{equation}
\frac{d}{dt} P_{t}(L=\infty) = 
-\frac{ P_t(L=\infty) }{t+s+1} + \frac{t+s}{(t+s+1)^2}.
\label{eq:dPtLinf}
\end{equation}

\noindent
The solution of Eq. (\ref{eq:dPtLinf}) is

\begin{eqnarray}
P_t(L=\infty) 
&=&
\frac{s+1}{t+s+1} P_0(L=\infty)
+ \frac{t}{t+s+1}
\nonumber \\
&-& \frac{1}{t+s+1} \ln  \left( \frac{t+s+1}{s+1} \right).
\label{eq:PtLinf}
\end{eqnarray}

\noindent
For $t=0$ the last two terms on the right hand side of Eq. (\ref{eq:PtLinf})
vanish and $P_t(L=\infty)$ coincides with the corresponding probability for
the seed network, which is given by $P_0(L=\infty)$.
The complementary probability, $P_t(L < \infty)$, is given by

\begin{eqnarray}
P_t(L <\infty) 
&=&
\frac{s+1}{t+s+1} P_0(L < \infty)
\nonumber \\
&+& \frac{1}{t+s+1} \ln \left( \frac{t+s+1}{s+1} \right).
\label{eq:PtLinf2}
\end{eqnarray}

\noindent
In the long time limit, the first term of Eq. (\ref{eq:PtLinf2}) declines
faster than the second term and the effect of the seed network 
slowly vanishes. As $t \rightarrow \infty$ the probability
$P_t(L < \infty)$ converges to its asymptotic form

\begin{equation}
P_t(L < \infty) \rightarrow \frac{1}{t+s+1} \ln (t+s+1) \rightarrow \frac{1}{t} \ln t.
\end{equation}

Upon formation of the daughter node, it acquires an
outgoing link to the mother node and with
probability $p$ to each one of its outgoing neighbors.
Therefore, 

\begin{equation}
P_t^{\rm D}(L=1) = 
p 
P_t^{\rm M}(L=1) 
+ \frac{1}{N_t+1}.
\label{eq:Pd1}
\end{equation}

\noindent
Since the mother node, M, is a randomly selected node at time $t$,
one can replace the probability 
$P_t^{\rm M}(L=1)$ 
by
$P_t(L=1)$.
As a result, Eq.
(\ref{eq:Pd1}) 
is replaced by

\begin{equation}
P_t^{\rm D}(L=1) = 
p
P_t(L=1) 
+
\frac{1}{N_t+1}.
\label{eq:Pd1p}
\end{equation}

\noindent
The case of paths of length $\ell=1$ is special in the sense that 
there is no degeneracy. 
Therefore, the parameter that appears
in Eq. (\ref{eq:Pd1p})
and in the second term on the right hand side of Eq. 
(\ref{eq:PdellLeq2}) is $p$ rather than $\eta$.
However, the replacement of $p$ by $\eta$ in these two equations
greatly simplifies the analysis.
The resulting equations are

\begin{equation}
P_t^{\rm D}(L=1) = 
\eta
P_t(L=1) 
+
\frac{1}{N_t+1}.
\label{eq:Pd1eta}
\end{equation}

\noindent
and

\begin{equation}
P_t^{\rm D}(L=2) = \eta P_t^{\rm M}(L=2) + (1 - \eta) P_t^{\rm M}(L=1).
\label{eq:PdellLeq2eta}
\end{equation}

\noindent
These equations provide a very good approximation for $P_t(L=\ell)$.
This is due to a combination of two properties:
(a) the probability $P_t(L=1)$ is of order $1/N_t$; and (b)
The difference between $p$ and $\eta$ is of order $p^3(1-p)$, which
is small for most values of $p$.
For the sake or completeness, we present in Appendix C the exact
master equation, in which $p$ is not replaced by $\eta$.
In practice, it is found that the approximation gives rise to a slight
deviation in $P_t(L=1)$ and $P_t(L=2)$,
while the results for $P_t(L=\ell)$
are not affected in any noticeable way.

Incorporating the contribution of D to $P_{t+1}(L=1)$ we obtain

\begin{equation}
P_{t+1}(L=1) = 
\frac{  N_t(N_t-1) P_t(L=1) + N_t P_t^{\rm D}(L=1)  }
{N_t(N_t+1)}.
\label{eq:Ptinf}
\end{equation}

\noindent
Inserting $P_t^{\rm D}(L=1)$ from Eq. (\ref{eq:Pd1eta}),
we obtain

\begin{eqnarray}
P_{t+1}(L=1) 
&=& 
\frac{  N_t(N_t-1) P_t(L=1) + \eta N_t P_t(L=1) }
{N_t(N_t+1)} 
\nonumber \\
&+& \frac{1}{(N_t+1)^2}.
\label{eq:Ptinfpt}
\end{eqnarray}

\noindent
Subtracting $P_t(L=1)$ from both sides,
replacing the difference on the
left hand side by a time derivative
and replacing $N_t$ by $t+s$, we obtain

\begin{eqnarray}
\frac{d}{dt} P_t(L=1) 
&=&
- \left( \frac{2-\eta}{t+s+1} \right) P_t(L=1)
\nonumber \\
&+& \frac{1}{(t+s+1)^2}.
\label{eq:P1}
\end{eqnarray}

\noindent
The first term on the right hand side of Eq. (\ref{eq:P1})
accounts for the probabilistic links from D to outgoing neighbors of M,
while the second term accounts for the deterministic link from D to M.
The solution of Eq.
(\ref{eq:P1})
is given by

\begin{eqnarray}
P_t(L=1) 
&=&
 \frac{ (s+1)^{1-\eta} }{ (t+s+1)^{2-\eta} }
\left[ (s+1) P_0(L=1) - \frac{1}{1-\eta} \right]
\nonumber \\
&+&
\frac{1}{(1-\eta)(t+s+1)}.
\label{eq:P1s}
\end{eqnarray}

\noindent
At $t=0$ the probability $P_t(L=1)$ is reduced to $P_0(L=1)$.
In the long time limit

\begin{equation}
P_t(L=1) \rightarrow \frac{1}{(1-\eta)(t+s-1)} \rightarrow \frac{1}{(1-\eta)t}.
\end{equation}

For paths of lengths $\ell \ge 2$, the probability
$P_t^{\rm D}(L=\ell)$ is given by Eq. (\ref{eq:Pdell}).
Replacing $P_t^{\rm M}(L=\ell)$ by $P_t(L=\ell)$ we obtain

\begin{equation}
P_t^{\rm D}(L=\ell) = \eta P_t(L=\ell) + (1 - \eta) P_t(L=\ell-1).
\label{eq:Pdell2}
\end{equation}

\noindent
After the node duplication step is completed, the DSPL
at time $t+1$ is given by

\begin{equation}
P_{t+1}(L=\ell) =
\frac{N_t(N_t-1) P_t(L=\ell) 
+
N_t P_t^{\rm D}(L=\ell)}{N_t(N_t+1)}.
\label{eq:P_td1}
\end{equation}

\noindent
Inserting the expression for $P_t^{\rm D}(L=\ell)$ from Eq. (\ref{eq:Pdell2}),
we obtain

\begin{eqnarray}
P_{t+1}(L=\ell) 
&=&
\frac{ (N_t-1) P_t(L=\ell) 
+
\eta  P_t(L=\ell)
}
{N_t+1}
\nonumber \\
&+&
\frac{
(1-\eta)  P_t(L=\ell-1)}{N_t+1}.
\label{eq:P_td12}
\end{eqnarray}

\noindent
Subtracting $P_t(L=\ell)$ 
from both sides of Eq. (\ref{eq:P_td12}),
replacing the difference
on the left hand side
by a time derivative,
and replacing $N_t$ by $t+s$,
we obtain

\begin{eqnarray}
\frac{d}{dt} P_t(L=\ell) 
&=&
-  \left( \frac{ 2-\eta}{t+s+1} \right) P_t(L=\ell)
\nonumber \\ 
&+&
 \left( \frac{1-\eta}{t+s+1} \right) P_t(L=\ell-1)
\label{eq:Pell}
\end{eqnarray}

\noindent
where $\ell \ge 2$.
Summing up the equations for the time derivatives of
$P_t(L=\infty)$ [Eq. (\ref{eq:dPtLinf})],
$P_t(L=1)$ [Eq. (\ref{eq:P1})]
and
$P_t(L=\ell)$ for $\ell \ge 2$
[Eq. (\ref{eq:Pell})],
it is found that the right hand sides sum up to zero,
namely the normalization of the DSPL is maintained.

The solution of Eq. 
(\ref{eq:Pell}),
for $\ell \ge 2$,
is given by

\begin{eqnarray}
P_t(L=\ell) 
&=&
\frac{1}{t_s^{2-\eta}}
\sum_{\ell^{\prime}=1}^{ \min\{\ell,\Delta_0 \} } 
\frac{(1-\eta)^{\ell-\ell'}}{(\ell-\ell')!} 
(\ln  t_s)^{\ell - \ell'}
P_0(L=\ell^{\prime})
\nonumber \\
&+&
\frac{1}{(1-\eta)(s+1) t_s^{2-\eta}}
\sum_{\ell'=\ell}^{\infty}
\frac{(1-\eta)^{\ell'}}{\ell' !} 
(\ln t_s)^{\ell'},
\label{eq:Pells}
\end{eqnarray}

\noindent
where $\Delta_0$ is the diameter of the seed network and

\begin{equation}
t_s = \frac{t+s+1}{s+1}.
\end{equation}

\noindent
The first term in Eq. (\ref{eq:Pells}) accounts for the DSPL of the
seed network and for the directed paths that emerge between newly formed nodes 
to seed-network nodes. The second term in Eq. (\ref{eq:Pells}) accounts for the directed
paths that emerge between pairs of nodes that form during the growth phase of the network.
At $t=0$ the second term in Eq. (\ref{eq:Pells}) vanishes
and the distribution $P_t(L=\ell)$ is reduced to the DSPL of the
seed network, given by $P_0(L=\ell)$.
As the network grows, the first term in Eq. (\ref{eq:Pells}),
which captures the DSPL of the seed network, declines and
the second term becomes dominant.

In conclusion,
Eqs. 
(\ref{eq:PtLinf}),
(\ref{eq:P1s})
and
(\ref{eq:Pells})
provide a closed form expression for the DSPL of the directed corded DND network
at time $t$ for any size and degree distribution of the seed network.
Since not all pairs of nodes $i$ and $j$ are connected by directed paths
from $i$ to $j$,
it is convenient to introduce an adjusted form of the DSPL,
which accounts only for the pairs of nodes which are connected
by directed paths. It is denoted by

\begin{equation}
P_t^{\cal C}(L=\ell) = \frac{P_t(L=\ell)}{P_t(L < \infty)}.
\end{equation}

\begin{figure}
\begin{center}
\includegraphics[width=6.0cm]{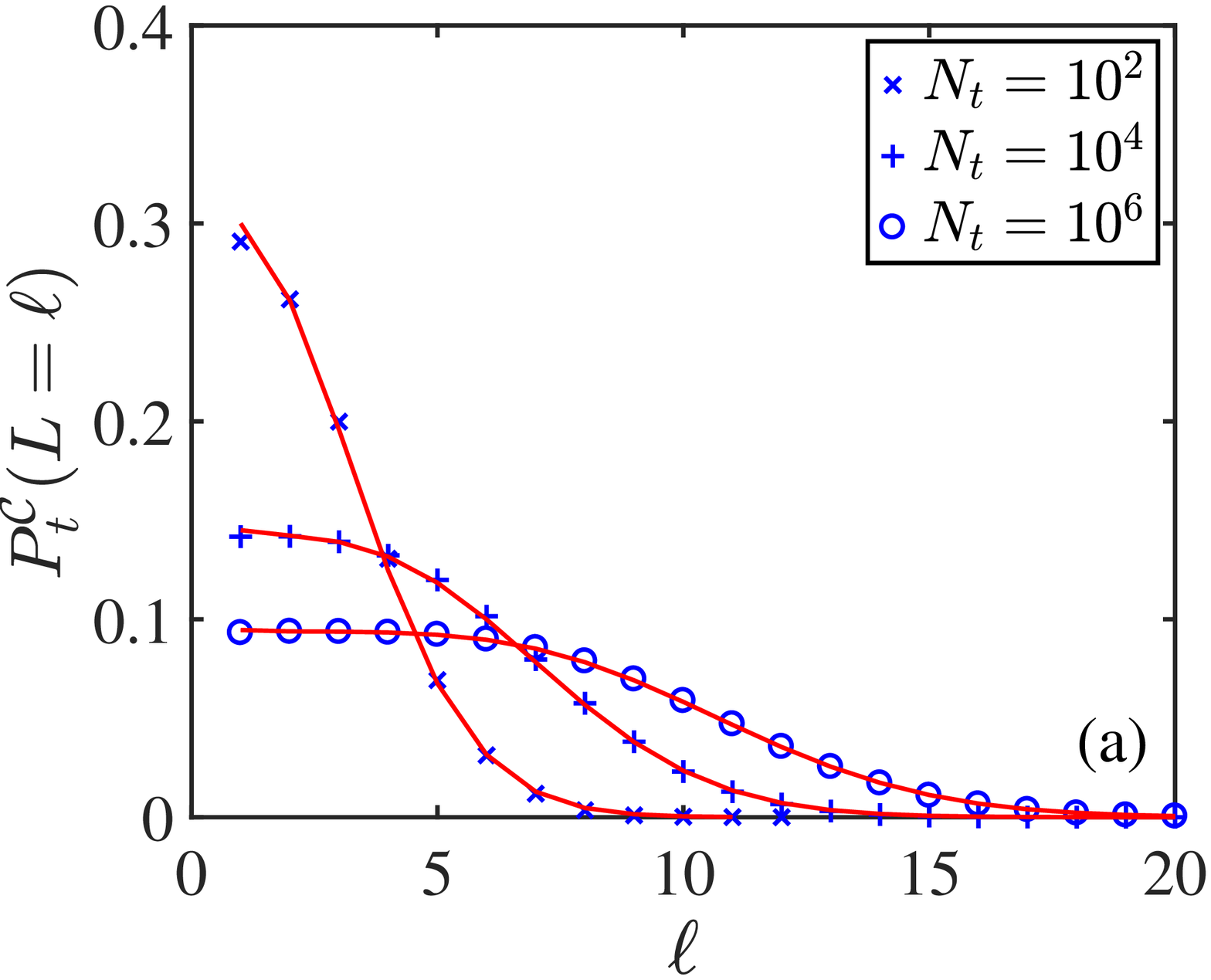} \\
\includegraphics[width=6.0cm]{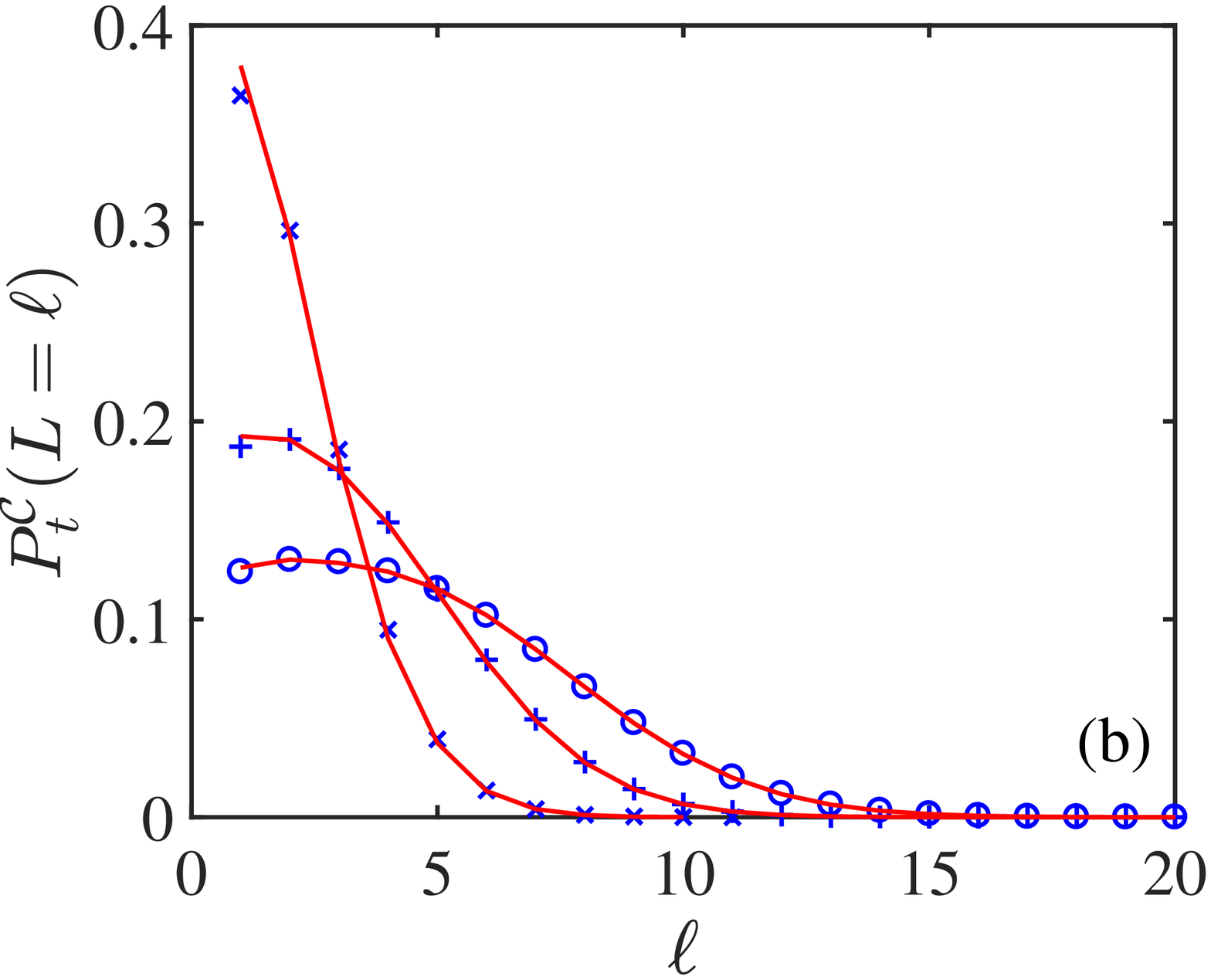} \\
\includegraphics[width=6.0cm]{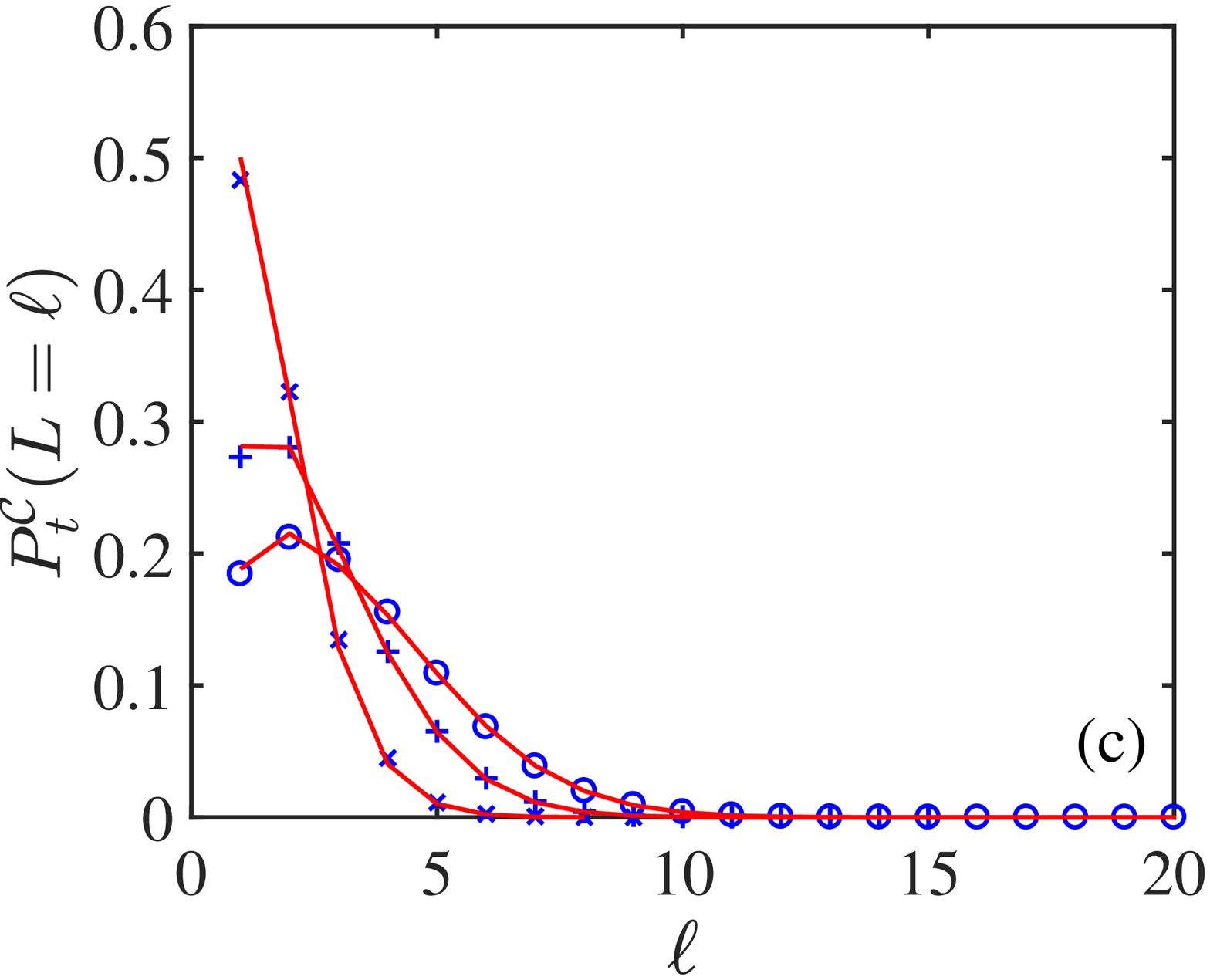} \\
\includegraphics[width=6.0cm]{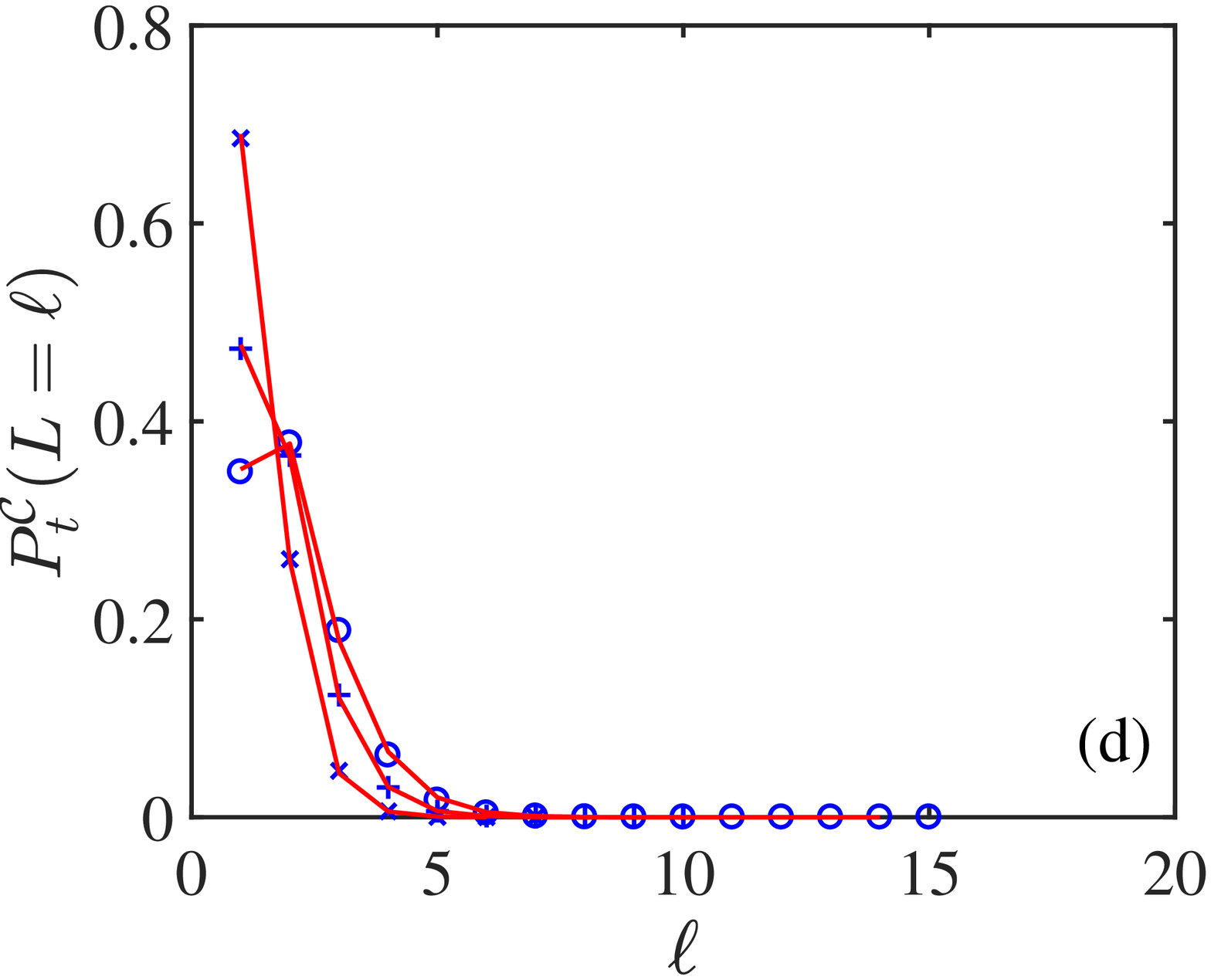} \\
\caption{
Analytical results (solid lines) for the
distribution $P_t^{\cal C}(L=\ell)$
of the corded DND network with 
(a) $p=0.2$; 
(b) $p=0.4$;
(c) $p=0.6$;
and
(d) $p=0.8$,
for network sizes of
$N_t=10^2$, $10^4$ and $10^6$.
The analytical results
are found to be in very good agreement with the results of computer simulations
(symbols),
obtained by averaging over $100$ network instances.
As $p$ is increased, the distances become shorter and the DSPL becomes narrower.
}
\label{fig:6}
\end{center}
\end{figure}

\noindent
In Fig. 
\ref{fig:6}
we present the adjusted DSPL, 
denoted by $P_t^{\cal C}(L=\ell)$
vs. $\ell$ 
for an ensemble of corded DND networks of sizes 
$N_t=10^2$, $10^4$ and $10^6$,
grown from a seed network of size $s=2$,
with
$p=0.2, 0.4, 0.6$ and $0.8$.

\section{The mean distance}

The mean distance between pairs of nodes in the corded DND networks
is obtained by averaging the path lengths among pairs
of nodes $i$ and $j$, that are connected by directed paths from $i$ to $j$.
It is given by

\begin{equation}
{\mathbb E}_t[L|L<\infty] = \sum_{\ell \ge 1}
\ell  P_t^{\cal C}(L=\ell).
\label{eq:<d>tdef}
\end{equation}

\noindent
It can also be expressed in the form

\begin{equation}
{\mathbb E}_t[L|L<\infty] = 
\frac{ \sum\limits_{\ell \ge 1} \ell  P_t (L=\ell)}
{P_t(L < \infty)}.
\label{eq:<d>tdef2}
\end{equation}

Carrying up the summation in the numerator of Eq. 
(\ref{eq:<d>tdef2})
we obtain

\begin{eqnarray}
{\mathbb E}_t[L|L<\infty] &=&
 \frac{ (s+1) P_0(L<\infty) {\mathbb E}_0[L|L<\infty]   } { (s+1)P_0(L<\infty) + \ln t_s } 
\nonumber \\
&+&
 \frac{  \left[ 1 +  (1-\eta) (s+1)P_0(L<\infty) \right]   \ln t_s }{ (s+1)P_0(L<\infty) + \ln t_s } 
\nonumber \\
&+&
 \frac{(1-\eta) (\ln t_s)^2 }{ 2[(s+1)P_0(L<\infty) + \ln t_s] }  
\label{eq:EL1}
\end{eqnarray}

\noindent
The first term on the right hand side of 
Eq. (\ref{eq:EL1}) accounts for the contribution of paths
between pairs of nodes that reside on the seed network. 
The second term
accounts for paths from nodes that formed during the growth phase  
to nodes in the seed network, while the third term accounts for paths between
pairs of nodes that formed during growth.
At $t=0$ the rescaled time is $t_s=1$ and $\ln t_s=0$.
As a result, the initial value of ${\mathbb E}_t[L|L<\infty]$ is indeed
${\mathbb E}_0[L|L<\infty]$.
The first term in Eq. (\ref{eq:EL1}) decreases monotonically as time proceeds,
the second term initially increases and then starts to decrease, 
while the third term increases monotonically in time.
In the long time limit the third term becomes dominant and

\begin{equation}
{\mathbb E}_t[L|L<\infty] \rightarrow \frac{1-\eta}{2} \ln t.
\end{equation}

\begin{figure}
\begin{center}
\includegraphics[width=7cm]{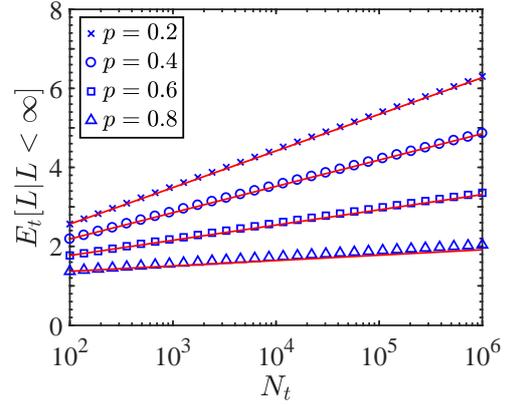}
\caption{
Analytical results (solid lines) for the
mean shortest path length  
${\mathbb E}_t[L|L<\infty]$, 
of the corded DND network,
as a function of network size $N_t$, 
obtained from
Eq. (\ref{eq:EL1}),
for $p=0.2$, $0.4$, $0.6$ and $0.8$.
The analytical results are in very good agreement with the simulation results
(symbols),
confirming the logarithmic dependence of 
${\mathbb E}_t[L|L<\infty]$  
on the network size.
As $p$ is increased, the network becomes more dense and the
slope of
${\mathbb E}_t[L|L<\infty]$ 
as a function of $\ln N_t$ decreases.
For clarity, we focus on network sizes in the range
$10^2 \le N_t \le 10^6$.
}
\label{fig:7}
\end{center}
\end{figure}

In Fig. \ref{fig:7} we present the mean distance, ${\mathbb E}_t[L|L<\infty]$,
as a function of the network size $N_t$, for $p=0.2, 0.4, 0.6$ and $0.8$.
The analytical results, obtained from Eq. (\ref{eq:EL1}),
where $\eta$ is taken from Eq.
(\ref{eq:eta5}),
are found to be in very good agreement with 
the results obtained from computer simulations (symbols).

\section{The variance of the DSPL}

In order to obtain the variance of the DSPL, 
we need to calculate its second moment, which is given by

\begin{equation}
{\mathbb E}_t[L^2|L<\infty]
= \sum_{\ell \ge 1}  \ell^2 P_t^{\cal C}(L=\ell).
\end{equation}

\noindent
Carrying out the summations, we obtain

\begin{eqnarray}
{\mathbb E}_t[L^2|L<\infty] &=& 
 \frac{ 
(s+1) P_0(L<\infty) {\mathbb E}_0[L^2|L<\infty] }
{(s+1) P_0(L<\infty)
 + 
\ln t_s } 
\nonumber \\
&+&
\frac{ \left[1 + (1-\eta) P_0(L<\infty) \right] \ln t_s  }
{(s+1) P_0(L<\infty)
 + 
\ln t_s     } 
\nonumber \\
&+&
 \frac{   2 (1-\eta) P_0(L<\infty)  {\mathbb E}_0[L|L<\infty]     \ln t_s  }
{(s+1) P_0(L<\infty)
 + 
\ln t_s     } 
\nonumber \\
&+&
 \frac{
(1-\eta) ( 5-2\eta )  P_0(L<\infty)   (\ln t_s)^2
   }
{2[(s+1) P_0(L<\infty)
 + 
\ln t_s  ]   } 
\nonumber \\
&+&
\frac{ (1-\eta)^2 (\ln t_s)^3 }
{3[ (s+1)P_0(L<\infty) + \ln t_s ]} .
\label{eq:EL2}
\end{eqnarray}

\noindent
The first term on the right hand side of Eq. (\ref{eq:EL2}) 
accounts for paths  between pairs of nodes
that reside on the seed network. 
The second and third terms account for paths from nodes
that form during growth and nodes in the seed network, while the last two terms
account for paths between pairs of nodes that form during growth.
Clearly, the initial value of the second moment at $t=0$ is
given by ${\mathbb E}_0[L^2|L<\infty]$.
As time proceeds the first term decreases monotonically, the second and third terms
initially increase and then start to decrease, while the last two terms
increase monotonically.
In the long time limit

\begin{equation}
{\mathbb E}_t[L^2|L<\infty] \rightarrow \frac{1}{3} (1-\eta)^2 (\ln t)^2.
\end{equation}

\noindent
The variance of $P_t(L=\ell)$ is given by

\begin{equation}
{\rm Var}_t(L) = {\mathbb E}_t[L^2|L<\infty]^2 - {\mathbb E}_t[L|L<\infty]^2,
\label{eq:VarL}
\end{equation}

\noindent
where ${\mathbb E}[L^2|L<\infty]$ 
is given by Eq. (\ref{eq:EL2}) and
${\mathbb E}[L|L<\infty]$ 
is given by Eq. (\ref{eq:EL1}).
In the long time limit, the variance converges to

\begin{equation}
{\rm Var}_t(L) = \frac{1}{12} (1-\eta)^2 (\ln t)^2.
\end{equation}

\begin{figure}
\begin{center}
\includegraphics[width=7cm]{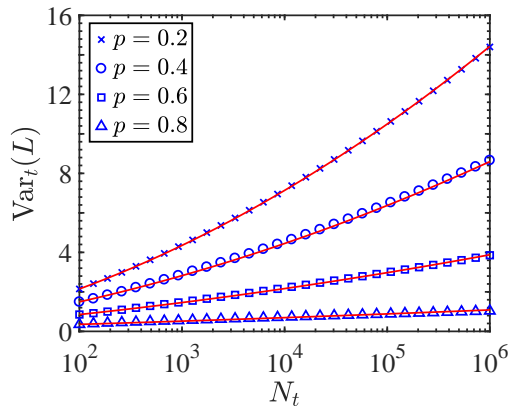}
\caption{
Analytical results (solid lines) for the variance ${\rm Var}_t(L)$, of
the distribution $P_t^{\cal C}(L=\ell)$
of the corded DND network,
as a function of network size $N_t$, 
obtained from
Eq. (\ref{eq:VarL})
for $p=0.2$, $0.4$, $0.6$ and $0.8$.
The analytical results are in very good agreement with the simulation results
(symbols).
}
\label{fig:8}
\end{center}
\end{figure}

In Fig. \ref{fig:8} we present the 
variance, ${\rm Var}_t(L)$,
of the DSPL
of the corded DND network as a function of network size, $N_t$,
for $p=0.2, 0.4, 0.6$ and $0.8$.. 
The analytical results (solid lines),
obtained from Eq.
(\ref{eq:VarL}),
where $\eta$ is taken from Eq.
(\ref{eq:eta5}),
are found to be in good agreement with the results of numerical
simulations (symbols),
thus the logarithmic scaling is confirmed.

\section{Discussion}

In an earlier paper we studied the DSPL of a corded ND network model
in which the links are undirected. 
Starting from a single-component seed network, the
network maintains its single component structure.
In this case any pair of nodes is connected by at least one path.
We obtained exact analytical results for the DSPL of this network, which consists
of two terms. The first term depends on the DSPL of the seed network and
dominates the results at short times, while the
second term is independent of the DSPL of the seed network and 
dominates the results at long times.
It was found that the mean distance is given by
$\langle L \rangle_t \simeq 2(1-\eta) \ln N_t$,
which means that the resulting network is a small world network.
The undirected ND network undergoes a phase transition at $p=1/2$,
above which the network becomes extremely dense
\cite{Lambiotte2016,Bhat2016}.
As a result, for $p \rightarrow 1/2$ from below, the shortest paths become
highly degenerate, in the sense that most pairs of nodes are connected by
several shortest paths, which all have the same length.

The corded DND model differs from its undirected counterpart is several ways.
Unlike the undirected model in which D may form probabilistic links to all the neighbors
of M, in the directed model it may form (directed) probabilistic links only to the outgoing
neighbors of M and not to the incoming neighbors of M.
In the undirected network each pair of nodes is connected by at least
one path. In contrast, from each node in the directed network one can 
access only older nodes which reside along the branch of the backbone
tree which leads to the sink node.
As a result, the probability that a random pair of nodes are connected
by a directed path is $P_t(L<\infty) \simeq \ln N_t / N_t$.
However, the mean distance between pairs of nodes which are connected
by directed paths exhibits the same qualitative behavior as in the undirected
network, namely ${\mathbb E}_t[L|L<\infty] \simeq [(1-\eta)/2] \ln N_t$.

The corded DND model may be useful in the study of scientific
citation networks. In these networks each new paper emanates
from one or more papers which were previously published in
the literature. The earlier papers, which are cited in the new paper,
are analogous to the mother node in the corded DND network.
More specifically,
each paper is represented by a node and each citation
is represented by a directed link from the citing paper to the cited paper.
The in-degree of each node is the number of citations received by the
corresponding paper, while the out-degree is the number of papers 
that appear in the reference list at the end of the paper.
Clearly, the out-degree of a paper is easily accessible and is 
fixed once the paper is published.
In contrast, the in-degree of a paper is initially 
zero and it may grow as the paper gets
cited by subsequent papers.
The citations of each paper are spread throughout the scientific literature.
Gathering this information requires an effort.
It can be obtained from search engines such as the 
Institute of Scientific Information (ISI) Web of Knowledge
and Google Scholar.

The corded DND model captures some essential properties 
of scientific citation networks.
It is a directed network whose links point 
from the citing paper to the cited paper.
The probabilistic links from the daughter node 
to outgoing neighbors of the mother node
correspond to the fact that a citation 
of a paper is often accompanied by citations to some
of the earlier papers that appear in its reference list. 
These probabilistic links also invoke
the preferential attachment mechanism, 
because the probability of a node to receive
such link is proportional to its in-degree. 
This is the mechanism that gives rise to the
power-law tail of the in-degree distribution.
The shift in the power-law degree distribution 
is due to the fact that the deterministic
links are formed by random attachment 
with no preference to high degree nodes.

The corded DND network provides some insight on the structure
of the scientific citation networks. In particular, it indicates
that for a given paper, the typical number of papers that are connected
to it by directed paths of citations 
(in the past or future)
scales like $\ln N/N$, where
$N$ is the network size.
This implies that the scientific literature is highly fragmented
in the sense that most pairs of papers are not connected via 
chains of citations and thus the network is not a small-world network.
For those pairs of papers that are connected by chains of
subsequent citations, the DSPL provides the 
breakdown into direct citations, indirect citations via a
single intermediate paper and indirect citations of higher 
orders.
This sheds new light on the way the impact of a paper
may be evaluated, namely not only in terms of the direct
citations but also in terms of the cumulative effect of
all the secondary citations.
Another aspect revealed by the model is that the structure
of citation networks evolves slowly, with a typical time scale
which is logarithmic in the network size.
This is in spite of the random nature of the growth process.
However, it should be emphasized that the corded DND model
should be considered as a minimal model of citation networks.
In more complete models a new paper may cite several 'mother nodes' 
as well as some of the earlier papers cited in them.
This would increase the number of directed paths but is not
expected to change the qualitative properties of the network.

\section{Summary}

We obtained exact analytical results for the time dependent
DSPL in a model of directed network 
that grows by a node duplication mechanism.
In this model, at each time step a random mother node is 
duplicated.
The daughter node acquires a directed link to the mother node,
and with probability $p$ it acquires a directed link to each
one of the outgoing neighbors of the mother node.
To obtain the DSPL we
derived a master equation for the
time evolution of the probability $P_t(L=\ell)$.
Finding an exact analytical solution of the master equation, we
obtained a closed form expression for the DSPL, in which
the probability $P_t(L=\ell)$
is expressed as a sum of two terms.
The first term is a convolution between the
DSPL of the seed network,
$P_0(L=\ell)$,
and a Poisson distribution.
The second term is a 
sum of Poisson distributions.
The expression for $P_t(L=\ell)$ is valid at 
all times and is not merely an asymptotic result.
We calculated the mean distance 
between pairs of nodes that are connected by directed paths
and showed that in the long time limit it scales like
${\mathbb E}_t[L|L<\infty] \simeq \frac{1-\eta}{2} \ln N_t$.
Thus, the mean distance between pairs of nodes that
are connected by directed paths scales logarithmically with the 
network size, as in small-world networks.
However, the fraction of pairs of nodes that are connected by directed
paths diminishes like $\ln N_t/N_t$ as the network size increases,
while most pairs of nodes are not connected by directed paths.
Therefore, the corded DND network is not a small-world network,
unlike the corded undirected node duplication network.

\vspace{0.1in}

\noindent
{\large \bf Acknowledgements}

This work was supported by the Israel Science Foundation grant no. 1682/18.

\vspace{0.1in}

\noindent
{\large \bf Author contribution statement}

All three authors contributed at all stages of the project.

\appendix

\vspace{0.1in}

\numberwithin{equation}{section}

\section{Canonical set of configurations and the degeneracies of shortest paths}

\begin{figure}
\begin{center}
\includegraphics[width=7cm]{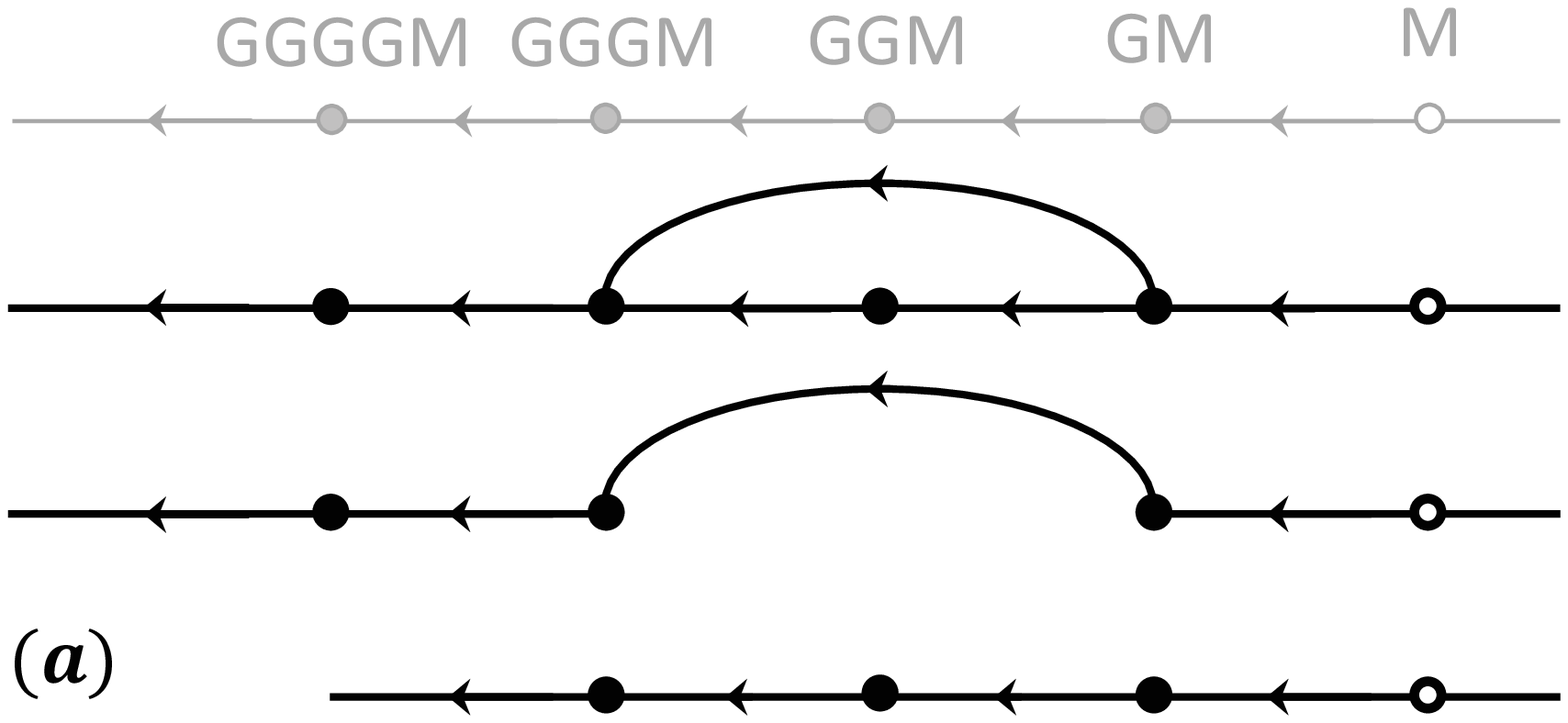}
\vspace{0.5in}
\\
\includegraphics[width=7cm]{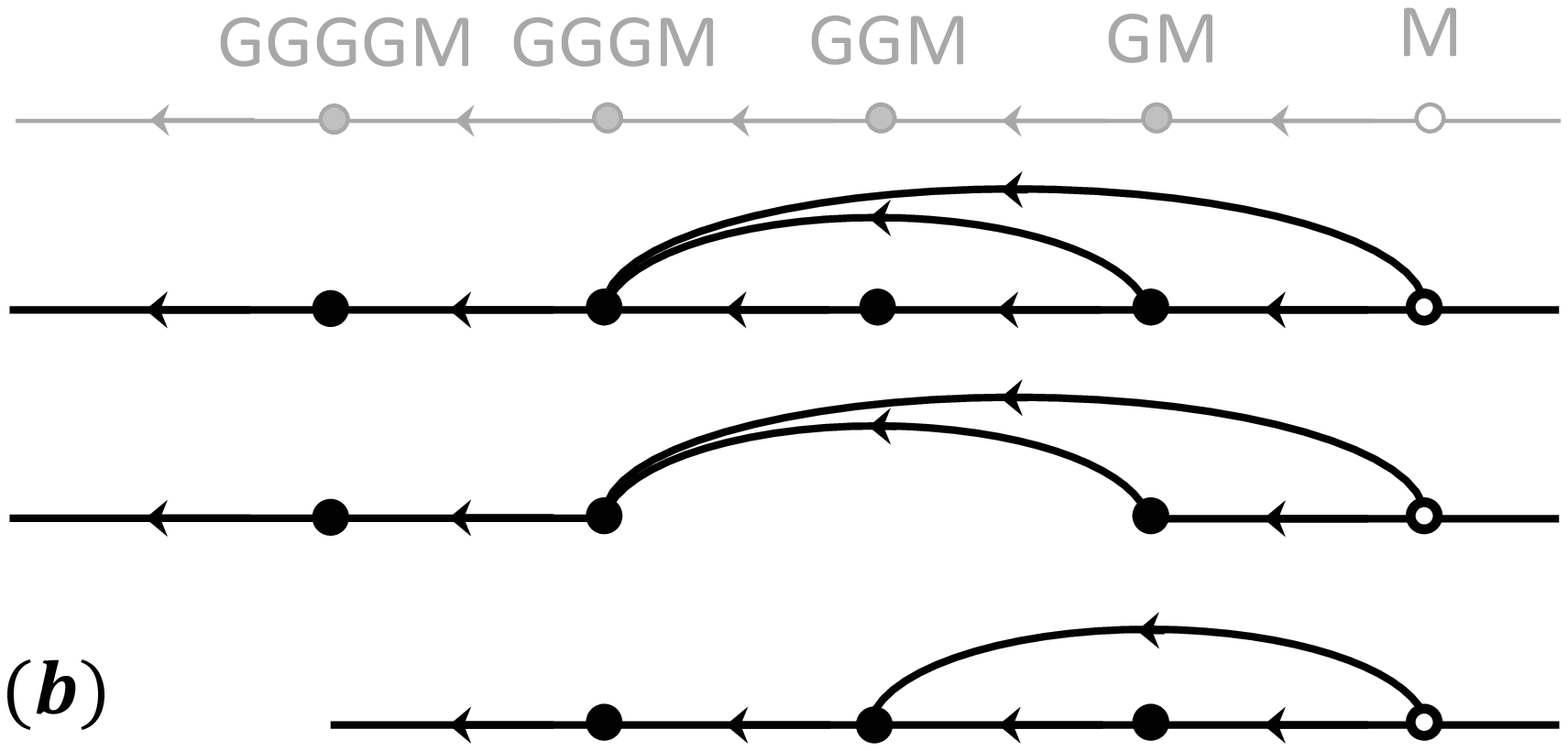}
\caption{
(a) The configuration
obtained upon duplication of a node M of configuration $g_{\rm M}=1^{**}$,
in which neither of the two outgoing links of M was duplicated. The degeneracy
of the shortest paths from D to downstream nodes is $g=1$.
Due to the shortcut from M to GGM, none of these shortest paths goes 
through GM. 
Moreover, GM is not directly connected to D and thus it cannot gain
new incoming links upon duplication of D.
Thus, the node GM has no effect on the degeneracy of the shortest paths
from D (and its descendants) to downstream nodes. 
Thus, its deletion has no effect on these shortest paths and their degeneracies.
The deletion of the node GM results in a canonical configuration of
the form $g_{\rm D}=1^{*}$;
(b) A configuration obtained from the
duplication of a node M for which $g_{\rm M}=1^{**}$, where
the link to GGM was duplicated while the link to GM
was not duplicated.
In this case none of the shortest paths from D to upstream nodes go 
through GM and it can thus be deleted, giving rise to the canonical
configuration $g_{\rm D}=1^{**}$.
}
\label{fig:9}
\end{center}
\end{figure}

In the master equation,
$\vec P_t^{\rm D}(G=g) = T \vec P_t^{\rm M}(G=g)$,
that describes the evolution of the degeneracies of shortest paths
from mother to daughter nodes, we use a canonical set of downstream
configurations. The low level configurations in this set are shown in
Fig. \ref{fig:3}. 
In addition to the canonical configurations, the duplication step may
lead to configurations which are not included in the canonical set.
Each one of these configurations is equivalent to a specific canonical
configuration in the sense that they both exhibit the same level of degeneracy
of the shortest paths from the newly formed node, D, to upstream nodes.
To illustrate this property, we present in Fig. \ref{fig:9}(a) the configuration
obtained upon duplication of a node, M, of configuration $g=1^{**}$,
in which neither of the two outgoing links of M was duplicated. The degeneracy
of the shortest paths from D to downstream nodes is $g=1$.
Due to the shortcut from M to GGM, none of these shortest paths goes 
through GM. Therefore, the betweeness centrality of GM is $0$.
Moreover, GM is not directly connected to D and thus it cannot gain
new incoming links upon duplication of D.
Thus, the node GM has no effect on the degeneracy of the shortest paths
from D (and its descendants) to downstream nodes. It can thus be deleted,
as shown in Fig. \ref{fig:9}(a), giving rise to a canonical configuration of
the form $g=1^{*}$.
In Fig. \ref{fig:9}(b) we present a configuration obtained from the
duplication of a node M for which $g_{\rm M}=1^{**}$.
In this case, the link to GGM was duplicated while the link to GM
was not duplicated.
In this case none of the shortest paths from D to upstream nodes go 
through GM and it can thus be deleted, giving rise to the canonical
configuration $g_{\rm D}=1^{**}$.

\section{Master equation for the degeneracies}

In order to analyze the temporal evolution of the degeneracies, we derive below the
master equation for the distribution of degeneracy levels. For simplicity, we consider
the case of single and double degeneracies, in which the distribution
$\vec P_t(G=g)$ is given by
Eq. (\ref{eq:3st})
and the transition matrix, $T$ is given by Eq. (\ref{eq:T33}).
The degeneracy distribution of the daughter node, which
is formed at time $t$, is given by

\begin{eqnarray}
P_t^{\rm D}(G=1^{*}) &=& 
(1-p) P_t^{\rm M}(G=1^{*}) 
\nonumber \\
&+& (1-p)^2 P_t^{\rm M}(G=1^{**}) 
\nonumber \\
&+& (1-p)^2 P_t^{\rm M}(G=2)
\nonumber \\
P_t^{\rm D}(G=1^{**}) &=& 
p P_t^{\rm M}(G=1^{*}) 
+ p P_t^{\rm M}(G=1^{**})
\nonumber \\ 
&+& 2p(1-p) P_t^{\rm M}(G=2)
\nonumber \\
P_t^{\rm D}(G=2) &=& 
p(1-p) P_t^{\rm M}(G=1^{**}) 
\nonumber \\
&+& p^2 P_t^{\rm M}(G=2)
\label{eq:PtDA}
\end{eqnarray}

\noindent
The degeneracy distribution at time $t+1$ is given by

\begin{equation}
\vec P_{t+1}(G=g) = \frac{ N_t \vec P_t(G=g) + \vec P_t^{\rm D}(G=g) }{N_t+1}, 
\label{eq:Ptp1GA}
\end{equation}

\noindent
where $g=1^{*}, 1^{**}$ and $2$.
Inserting $\vec P_t^{\rm D}(G=g)$ from Eq. (\ref{eq:PtDA}) into Eq. (\ref{eq:Ptp1GA}),
subtracting $\vec P_t(G=g)$ from both sides and replacing the difference on the left hand
side by a time derivative, we obtain

\begin{eqnarray}
\frac{d}{dt} P_t(G=1^{*}) &=& \frac{ 
- p P_t(G=1^{*})  }
{N_t+1}
\nonumber \\
&+& \frac{ (1-p)^2 [ P_t(G=1^{**}) +  P_t(G=2) ] }
{N_t+1}
\nonumber \\
\frac{d}{dt} P_t(G=1^{**}) 
&=& \frac{   p P_t(G=1^{*}) - (1-p)  P_t(G=1^{**}) }{N_t+1}
\nonumber \\
&+& \frac{ 2p(1-p) P_t(G=2)  }{N_t+1}
\nonumber \\
\frac{d}{dt} P_t(G=2) 
&=& \frac{  p(1-p) P_t(G=1^{**}) }{N_t+1}
\nonumber \\
&-& \frac{ (1-p)^2 P_t(G=2) }{N_t+1}.
\end{eqnarray}

\noindent
The solution of this master equation is given by

\begin{eqnarray}
P_t(G=1^{*}) &=& 
\frac{(1-p)^2}{1-p+p^2}
+ \frac{ P_0(G=1^{*}) }{t_s^{1-p+p^2}} 
\nonumber \\
&-& \frac{(1-p)^2}{(1-p+p^2) t_s^{1-p+p^2} }
\nonumber \\
P_t(G=1^{**}) &=&
\frac{p(1+p)}{(1+2p)(1-p+p^2)}
\nonumber \\
&-& 
  \frac{(1-2p) P_0(G=1^{*}) }{(2-3p) t_s^{1-p+p^2}}
\nonumber \\
&+& \frac{ (1-2p)(1-p)^2 }{(2-3p)(1-p+p^2)t_s^{1-p+p^2}}  
\nonumber \\
&+&
    \frac{(1-2p)P_0(G=1^{*})}{(2-3p) t_s^{1+p-2p^2}}
+ \frac{P_0(G=1^{**})}{ t_s^{1+p-2p^2} }
\nonumber \\
&-& \frac{1+p-4p^2}{ (1+2p) (2-3p) t_s^{1+p-2p^2}}
\nonumber \\
P_t(G=2) &=&
\frac{p^2}{(1+2p)(1-p+p^2)}
\nonumber \\
&-& 
\frac{ (1-p) P_0(G=1^{*}) }{ (2-3p) t_s^{1-p+p^2} } 
\nonumber \\
&-& \frac{(1-p)^2}{ (1-p+p^2) t_s^{1-p+p^2} }
\nonumber \\
&+&
  \frac{1+p-4p^2}{(1+2p)(2-3p)t_s^{1+p-2p^2}} 
\nonumber \\
&-& \frac{(1-2p) P_0(G=1^{*})  }{(2-3p) t_s^{1+p-2p^2} }
\nonumber \\
&-&  \frac{ P_0(G=1^{**}) }{t_s^{1+p-2p^2}}.
\end{eqnarray}

\noindent
The convergence of the degeneracy distribution,
$P_t(G=g)$ to its asymptotic form
follows a power-law function of the time with the two exponents:

\begin{eqnarray}
\alpha_1 &=& 1-p+p^2
\nonumber \\
\alpha_2 &=& 1+p-2p^2.
\end{eqnarray}

\noindent
The degeneracy distribution at steady state conditions
is given by

\begin{eqnarray}
P(G=1^{*}) &=& \frac{(1-p)^2}{1-p+p^2}
\nonumber \\
P(G=1^{**}) &=& \frac{p(1+p)}{(1+2p)(1-p+p^2)}
\nonumber \\
P(G=2) &=& \frac{p^2}{(1+2p)(1-p+p^2)}.
\end{eqnarray}

\noindent
The convergence rate to steady state is
dominated by the smaller exponent among $\alpha_1$ and $\alpha_2$,
namely $\alpha_{\rm min} = \min\{\alpha_1,\alpha_2\}$.
One can show that for $p < 2/3$, $\alpha_{\rm min}=\alpha_1$,
while for $p > 2/3$, $\alpha_{\rm min}=\alpha_2$.
The time its takes for the difference
$\Delta P_t(G=g) = P_t(G=g) - P(G=g)$
to go down to $1/e$ of its initial value is given by

\begin{equation}
\tau = (s+1) \left[ e^{1/\alpha_{\rm min}}-1 \right].
\end{equation}

\noindent
In the limit of $p \rightarrow 1^{-}$ the exponent $\alpha_2$
satisfies $\alpha_2 \rightarrow 0$.
As a result, the convergence slows down as $p$ is increased
towards $p=1$.

\section{Exact form of the master equation}

In the derivation of the master equation for $P_t(L=\ell)$, where
$\ell=1,2$, we used an approximation in which we
replaced the probability $p$ by $\eta$ in Eqs.
(\ref{eq:Pd1eta}) and (\ref{eq:PdellLeq2eta}).
Here we present the exact master equation, obtained from
Eqs. (\ref{eq:Pd1}) and (\ref{eq:Pdell}), in case that $p$ is
not replaced by $\eta$.
In this case, the equation for $P_t(L=1)$ takes the form

\begin{eqnarray}
\frac{d}{dt} P_t(L=1) 
&=&
- \left( \frac{2-p}{t+s+1} \right) P_t(L=1)
\nonumber \\
&+& \frac{1}{(t+s+1)^2},
\label{eq:P1p}
\end{eqnarray}

\noindent
which is similar to Eq. (\ref{eq:P1}) except for the replacement of $\eta$ by $p$.
The equation for $P_t(L=2)$ takes the form

\begin{eqnarray}
\frac{d}{dt} P_t(L=2) 
&=&
-  \left( \frac{ 2-\eta}{t+s+1} \right) P_t(L=2) 
\nonumber \\
&+&
 \left( \frac{1-p}{t+s+1} \right) P_t(L=1),
\label{eq:Pellp}
\end{eqnarray}

\noindent
which is similar to Eq. (\ref{eq:Pell}),
except for the replacement of $\eta$ by $p$ in the second term on the right hand side.
The time derivatives of $P_t(L=\ell)$ for $\ell \ge 3$ are given by Eq. (\ref{eq:Pell})
and the time derivative of $P_t(L=\infty)$ is given by Eq. (\ref{eq:dPtLinf}).
The solution for $P_t(L=1)$ is given by

\begin{eqnarray}
P_t(L=1) 
&=&
\frac{ 1 }{ t_s^{2-p} }
\left[  P_0(L=1) - \frac{1}{(1-p)(s+1)} \right]
\nonumber \\
&+&
\frac{1}{(1-p)(s+1)t_s},
\label{eq:P1speta}
\end{eqnarray}

\noindent
which is similar to Eq. (\ref{eq:P1s}), except for the replacement of $\eta$ by $p$.
The solution for $P_t(L=\ell)$, $\ell \ge 2$ is given by

\begin{eqnarray}
P_t(L=\ell) 
&=&
\frac{(1-p)}{t_s^{2-\eta}}
\frac{ (1-\eta)^{\ell-2}  (\ln  t_s)^{\ell - 1}  }{(\ell-1)!} 
P_0(L=1)
\nonumber \\
&+&
\frac{1-p}{t_s^{2-\eta}}
\left( \frac{1-\eta}{p-\eta} \right)^{\ell-2}
\times
\nonumber \\
& & \ \ \ \ \ \  \sum_{\ell^{\prime}=\ell}^{ \infty } 
\frac{(p-\eta)^{\ell'-1}  (\ln  t_s)^{\ell'}  }{\ell' !} 
P_0(L=1) 
\nonumber \\
&+&
\frac{1}{t_s^{2-\eta}}
\sum_{\ell^{\prime}=2}^{ \min\{\ell,\Delta_0 \} } 
\frac{ [ (1-\eta) \ln  t_s]^{\ell - \ell'}  }{(\ell-\ell')!} 
P_0(L=\ell^{\prime})
\nonumber \\
&-&
\frac{1}{ t_s^{2-\eta} }
\left( \frac{1-\eta}{p-\eta} \right)^{\ell-2}
\sum_{\ell^{\prime}=\ell}^{ \infty } 
\frac{(p-\eta)^{\ell'-1} (\ln  t_s)^{\ell'}  }{(s+1) \ell' !} 
\nonumber \\
&+&
\frac{1}{ t_s^{2-\eta} }
\sum_{\ell'=\ell}^{\infty}
\frac{(1-\eta)^{\ell'-1} (\ln t_s)^{\ell'}  }{ (s+1) \ell' !}. 
\label{eq:Pellspeta}
\end{eqnarray}

\noindent
The first two terms on the right hand side of Eq. (\ref{eq:Pellspeta})
correspond to the first term of Eq. (\ref{eq:Pells}),
except that in the contribution of
$P_0(L=1)$ is multiplied by $(1-p)/(1-\eta)$.
These terms account for the components of $P_t(L=\ell)$ that
depend on the DSPL of the seed network.
The third term adjusts for the difference between $p$ and $\eta$
in the terms that depend on the DSPL of the seed network. 
The fourth term corresponds to the second term of Eq. (\ref{eq:Pells}),
which does not depend on the DSPL of the seed network.
The last term adjusts for the difference between $p$ and $\eta$
in the terms that do not depend on the DSPL of the seed network.
The results for $P_t(L=\infty)$ are given
by Eq. (\ref{eq:PtLinf}).


\begin{thebibliography}{10}


\bibitem{Albert2002}
R. Albert, A.-L. Barab\'asi, 
{\it Rev. Mod. Phys.} {\bf 74}, 47 (2002) 

\bibitem{Caldarelli2007}
G. Caldarelli, 
{\it Scale free networks: complex webs in nature and technology} 
(Oxford University Press, 2007) 

\bibitem{Havlin2010}
S. Havlin, R. Cohen,
{\it Complex Networks: Structure, Robustness and Function}
(Cambridge University Press, 2010) 

\bibitem{Newman2010}
M.E.J. Newman, 
{\it Networks: an Introduction} 
(Oxford University Press, 2010) 

\bibitem{Estrada2011b}
E. Estrada,
{\it The Structure of Complex Networks: Theory and Applications}
(Oxford University Press, 2011) 

\bibitem{Barrat2012}
A. Barrat, M. Barth\'elemy, A. Vespignani,
Dynamical Processes on Complex Networks
(Cambridge University Press, 2012) 


\bibitem{Milgram1967}
S. Milgram, 
{\it Psychology Today} {\bf 1}, 61 (1967) 

\bibitem{Watts1998}
D. Watts, S. Strogatz,
{\it Nature} {\bf 393}, 440 (1998).

\bibitem{Chung2002}
F. Chung, L. Lu, 
{\it Proc. Nat. Acad. Sci. USA} {\bf 99}, 15879 (2002) 


\bibitem{Chung2003}
F. Chung, L. Lu, 
{\it Internet Mathematics} {\bf 1}, 91 (2003) 



\bibitem{Barabasi1999}
A.-L. Barab\'asi, R. Albert, 
{\it Science} {\bf 286},  509  (1999) 

\bibitem{Jeong2000}
H. Jeong, B. Tombor, R. Albert, Z.N. Oltvai, A.-L. Barab\'asi, 
{\it Nature} {\bf 407}, 651 (2000) 

\bibitem{Krapivsky2000}
P.L. Krapivsky, S. Redner, F. Leyvraz, 
{\it Phys. Rev. Lett.} {\bf 85},  4629   (2000) 

\bibitem{Krapivsky2001}
P.L. Krapivsky, S. Redner, 
{\it Phys. Rev. E} {\bf 63}, 066123 (2001)

\bibitem{Vazquez2003}
A. V\'azquez, 
{\it Phys. Rev. E} {\bf 67}, 056104 (2003)



\bibitem{Cohen2003}
R. Cohen, S. Havlin, 
{\it Phys. Rev. Lett.} {\bf 90}, 058701 (2003)


\bibitem{Giot2003}
L. Giot et al.,
{\it Science} {\bf 302} 1727 (2003)


\bibitem{Maayan2005}
A. Ma\'ayan, S.L. Jenkins, S. Neves, A. Hasseldine, E. Grace, B. Dubin-Thaler, 
N.J. Eungdamrong, G. Weng, P.T. Ram, J.J. Rice, A. Kershenbaum, G.A. Stolovitzky, 
R.D. Blitzer, R. Iyengar,
{\it Science} {\bf 309}, 1078 (2005)


\bibitem{Dijkstra1959}
E.W. Dijkstra,
{\it Numerische Mathematik} {\bf 1}, 269 (1959)

\bibitem{Delling2009}
D. Delling, P. Sanders, D. Schultes, D. Wagner, 
Engineering route planning algorithms, in 
{\it Algorithmics of large and complex networks: design, 
analysis, and simulation}, 
J. Lerner, D. Wagner, and K.A. Zweig (Eds.), 
p. 117 (2009)






\bibitem{Satorras2015}
R. Pastor-Satorras, C. Castellano, P. Van Mieghem, A. Vespignani, 
{\it Rev. Mod. Phys.} {\bf 87}, 925 (2015)





\bibitem{Bollobas2001}
B. Bollobas, 
{\it Random Graphs, Second Edition}
(Academic Press, London, 2001)

\bibitem{Durrett2007}
R. Durrett,
{\it Random Graph Dynamics}
(Cambridge University Press, Cambridge, 2007)




\bibitem{Fronczak2004}
A. Fronczak, P. Fronczak, J.A. Holyst, 
{\it Phys. Rev. E} {\bf 70}, 056110 (2004)




\bibitem{Newman2001b}
M.E.J. Newman,
{\it Proc. Natl. Acad. Sci. USA} {\bf 98}, 404 (2001)


\bibitem{Hartmann2017}
A.K. Hartmann, M. M\'ezard,
{\it Phys. Rev. E} {\bf 97}, 032128 (2017)



\bibitem{Newman2001}
M.E.J. Newman, S.H. Strogatz, D.J. Watts,
{\it Phys. Rev. E} {\bf 64}, 026118 (2001)



\bibitem{Dorogotsev2003}
S.N. Dorogotsev, J.F.F. Mendes, A.N. Samukhin, 
{\it Nuclear Physics B} {\bf 653}, 307 (2003)

\bibitem{Blondel2007}
V.D. Blondel, J.-L. Guillaume, J.M. Hendrickx, R.M. Jungers, 
{\it Phys. Rev. E} {\bf 76}, 066101 (2007)


\bibitem{Hofstad2007}
R. van der Hofstad, G. Hooghiemstra, D. Znamenski, 
{\it Electronic Journal of Probability} 
{\bf 12}, 703 (2007)

\bibitem{Esker2008}
H. van der Esker, R. van der Hofstad, G. Hooghiemstra, 
{\it J. Stat. Phys.} {\bf 133}, 169 (2008)


\bibitem{Shao2008}
J. Shao, S. V. Buldyrev, R. Cohen, M. Kitsak, S. Havlin, H. E. Stanley, 
{\it Europhys. Lett.} {\bf 84}, 48004 (2008)

\bibitem{Shao2009}
J. Shao, S.V. Buldyrev, L.A. Braunstein, S. Havlin, H.E. Stanley, 
{\it Phys. Rev. E} {\bf 80}, 036105 (2009)


\bibitem{Katzav2015}
E. Katzav, M. Nitzan, D. ben-Avraham, P.L. Krapivsky, 
R. K\"uhn, N. Ross, O. Biham,
{\it EPL} {\bf 111}, 26006 (2015)


\bibitem{Erdos1959}
P. Erd{\H o}s, A. R\'{e}nyi, 
{\it Publicationes Mathematicae (Debrecen)} {\bf 6}, 290 (1959)


\bibitem{Erdos1960}
P. Erd{\H o}s, A. R\'{e}nyi, 
{\it Publ. Math. Inst. Hung. Acad. Sci.} {\bf 5}, 17 (1960)

\bibitem{Erdos1961}
P. Erd{\H o}s, A. R\'{e}nyi, 
{\it Bull. Inst. Int. Stat.} {\bf 38}, 343 (1961)



\bibitem{Nitzan2016}
M. Nitzan, E. Katzav, R. K\"uhn, O. Biham,
{\it Phys. Rev. E} {\bf 93}, 062309 (2016)



\bibitem{Melnik2016}
S. Melnik, J.P. Gleeson, 
arXiv:1604.05521 


\bibitem{Molloy1995}
{M. Molloy, B. Reed},
{\it Random Struct. Algorithms} {\bf 6}, 161 (1995)


\bibitem{Molloy1998}
{M. Molloy, B. Reed},
{\it Combinatorics, Probability and Computing} {\bf 7 }, 295 (1998)



\bibitem{Bhan2002}
A. Bhan, D.J. Galas, T.G. Dewey, 
{\it Bioinformatics} {\bf 18}, 1486  (2002)


\bibitem{Kim2002}
J. Kim, P.L. Krapivsky, B. Kahng, S. Redner, 
{\it Phys. Rev. E} {\bf 66}, 055101 (2002)

\bibitem{Chung2003b}
F. Chung, L. Lu, T.G. Dewey, D.J. Galas, 
{\it J. Comput. Biol.} {\bf 10}, 677 (2003)

\bibitem{Krapivsky2005}
P.L. Krapivsky, S. Redner, 
{\it Phys. Rev. E} {\bf 71}, 036118 (2005)

\bibitem{Ispolatov2005}
I. Ispolatov, P.L. Krapivsky, A. Yuryev, 
{\it Phys. Rev. E} {\bf 71}, 061911 (2005)

\bibitem{Ispolatov2005b}
I. Ispolatov, P.L. Krapivsky, I. Mazo, A. Yuryev, 
{\it New J. Phys.} {\bf 7}, 145 (2005)

\bibitem{Bebek2006}
G. Bebek,  P.  Berenbrink, C. Cooper, T. Friedetzky, J. Nadeau, S.C. Sahinalp,
{\it Theor. Comput. Sci.} {\bf 369},  239 (2006)

\bibitem{Li2013}
S. Li, K.P. Choi, T. Wu, 
{\it Theor. Comput. Sci.} {\bf 476},  94 (2013)




\bibitem{Lambiotte2016}
R. Lambiotte, P.L. Krapivsky, U. Bhat, S. Redner,
{\it Phys. Rev. Lett.} {\bf 117}, 218301 (2016)


\bibitem{Bhat2016}
U. Bhat, P.L. Krapivsky, R. Lambiotte, S. Redner,
{\it Phys. Rev. E.} {\bf 94}, 062302 (2016)




\bibitem{Toivonen2009}
R. Toivonen, L. Kovanen, M. Kivel\"a, J.-P. Onnela, J. Saram\"aki, K. Kaski, 
{\it Social Networks} {\bf 31}, 240 (2009)

\bibitem{Granovetter1973}
M. Granovetter, 
{\it American Journal of Sociology} {\bf 78}, 1360 (1973)




\bibitem{Milo2002}
R. Milo, S. Shen-Orr, S. Itzkovitz, N. Kashtan, D. Chklovskii, U. Alon,
{\it Science} {\bf 298}, 824 (2002)

\bibitem{Alon2006}
U. Alon,
{\it An Introduction to Systems Biology: Design Principles of Biological Circuits}
(Chapman and Hall/CRC, 2006)


\bibitem{Steinbock2017}
C. Steinbock, O. Biham, E. Katzav,
{\it Phys. Rev. E} {\bf 96}, 032301 (2017)



\bibitem{Ohno1970}
S. Ohno, 
{\it Evolution by Gene Duplication} (Springer-Verlag, New York, 1970)

\bibitem{Teichmann2004}
S.A. Teichmann, M.M. Babu, 
{\it Nature Genetics} {\bf 36},  492  (2004)



\bibitem{Redner1998}
S. Redner,  
{\it Eur. Phys. J. B} {\bf 4}, 131 (1998)

\bibitem{Redner2005}
S. Redner,
{\it Physics Today} {\bf 58}, 49 (2005)

\bibitem{Radicchi2008}
F. Radicchi, S. Fortunato, C. Castellano,  
{\it Proc. Natl. Acad. Sci. USA} {\bf 105}, 17268 (2008)




\bibitem{Golosovsky2012}
M. Golosovsky, S. Solomon,
{\it Phys. Rev. Lett.} {\bf 109}, 098701 (2012)

\bibitem{Golosovsky2017a}
M. Golosovsky and S. Solomon,
{\it Phys. Rev. E} {\bf 95}, 012324 (2017)

\bibitem{Golosovsky2017b}
M. Golosovsky,
{\it Phys. Rev. E} {\bf 96}, 032306 (2017)


\bibitem{Peterson2010}
G.J. Peterson, Steve Press\'e, K.A. Dill,
{\it Proc. Natl. Acad. Sci. USA} {\bf 107 }, 16023 (2010)


\bibitem{Steinbock2018}
C. Steinbock, O. Biham, E. Katzav,
{\it J. Stat. Mech.} 083403 (2019).


\bibitem{Smythe1995}
R.T. Smythe, H. Mahmoud,
{\it Theory Probab. Math. Statist.} {\bf 51}, 1 (1995)

\bibitem{Drmota1997}
M. Drmota, B. Gittenberger,
{\it Random Struct. Alg.} {\bf 10}, 421 (1997)

\bibitem{Drmota2005}
M. Drmota, H.-K. Hwang,
{\it Adv. Appl Probab.} {\bf 37}, 321 (2005)




\end{thebibliography}
\end{document}